\documentclass[preprint, 3p]{elsarticle}

\newtheorem{theorem}{Theorem}[section]
\newtheorem{lemma}[theorem]{Lemma}
\newenvironment{proof}{\par\noindent\textit{Proof.} \ }{\hfill$\blacksquare$\par}

\usepackage{setspace}
\usepackage{geometry}
\biboptions{sort&compress}
\usepackage{amsmath}
\usepackage{amsfonts}
\usepackage{hyperref}
\usepackage{amssymb}
\usepackage{graphicx}
\graphicspath{Figures/}
\usepackage{algorithm}
\usepackage{algpseudocode}
\usepackage[caption=false]{subfig}
\usepackage{xcolor, soul}

\usepackage{framed}
\colorlet{shadecolor}{blue!10}

\begin{document}

\title{Improved Stability Estimates and Flight Time Predictions Using Higher-Order Transverse Discontinuity Mapping in Hybrid Dynamical Systems}

\author[a,b]{Rohit Chawla}
\ead{rohit.chawla@ucd.ie}

\author[a,b]{Aasifa Rounak}
\ead{aasifa.rounak@ucd.ie}

\author[a,b]{Vikram Pakrashi\corref{cor1}}
\ead{vikram.pakrashi@ucd.ie}

\affiliation[a]{organization={UCD Centre for Mechanics, Dynamical Systems and Risk Laboratory, School of Mechanical and Materials Engineering, University College Dublin},
city={Dublin},
country={Ireland}}

%\affiliation[2]{organization={SFI MaREI Centre, University College Dublin}, city={Dublin},
%country={Ireland}}

\affiliation[b]{organization={UCD Energy Institute, University College Dublin}, city={Dublin},
country={Ireland}}

\cortext[cor1]{Corresponding author}

% For Journal of Nonlinear Dynamics

%\author[1,2,3]{\fnm{Rohit} \sur{Chawla}} \email{rohit.chawla@ucdconnect.ie} \equalcont{All authors contributed equally to this work}

%\author[1,2,3]{\fnm{Aasifa} \sur{Rounak}} \email{aasifa.rounak@ucd.ie} \equalcont{All authors contributed equally to this work}

%\author*[1,2,3]{\fnm{Vikram} \sur{Pakrashi} \email{vikram.pakrashi@ucd.ie}} \equalcont{All authors contributed equally to this work}

%\affil*[]{\orgdiv{}, \orgname{}, \orgaddress{ \street{}, \city{}, \postcode{}, \state{}, \country{}}}

%\affil[1]{\orgdiv{UCD Centre for Mechanics, Dynamical Systems and Risk Laboratory, School of Mechanical and Materials Engineering}, \orgname{University College Dublin}, \orgaddress{\city{Dublin}, \country{Ireland}}}

%\affil[2]{\orgdiv{SFI MaREI Centre}, \orgname{University College Dublin}, \orgaddress{\city{Dublin}, \country{Ireland}}}

%\affil[3]{\orgdiv{UCD Energy Institute}, \orgname{University College Dublin}, \orgaddress{\city{Dublin}, \country{Ireland}}}

%-----------------------------------------------------------------------------------------------------------------

\begin{abstract}
\noindent This article emphasizes on inconsistencies in the dynamical estimates obtained by first-order transverse discontinuity mapping (TDM) and direct numerical observations for hybrid dynamical systems. Pitfalls of locally linearizing hybrid nonlinear dynamical systems near discontinuity boundaries are demonstrated along with examples of how such linearization could lead to incorrect estimates of impact occurrences for transverse interactions with a rigid barrier. A higher-order TDM is proposed to overcome this shortcoming, allowing for better analytical estimation of impact occurrence times, state transitions, and, consequently, the evolution of trajectories. The difference in flight times of two closely initiated trajectories in the local neighbourhood of a discontinuity boundary is estimated up to $\mathcal{O}(2)$. The resulting quadratic equation implies that the orbits local to the impacting state, corresponding to a negative discriminant, won't reach the discontinuity boundary. Further, the $\mathcal{O}(2)$ correction terms to the analytical expression of the TDM ensure that the flight time estimates do not diverge for low-velocity impacts near grazing, thereby avoiding overestimation of the mapped state. A numerical method is subsequently developed to estimate a saltation matrix incorporating the proposed higher-order TDM to avoid incorrect impact occurrences. Modifications to the existing algorithms used to numerically quantify local stability, namely the Lyapunov spectra and Floquet multipliers, are proposed.
%ensuring that the state of stability is appropriately computed. 
Stability analyses using the proposed higher-order approach are carried out for representative cases of a hard impact oscillator and a pair impact oscillator, with results consistent with numerically obtained bifurcation diagrams.
\end{abstract}

\begin{keyword}
   Discontinuity mapping \sep Saltation matrix \sep Vibro-impact oscillators \sep Bifurcation analysis \sep Lyapunov exponents \sep Floquet analysis
\end{keyword}

%\sep Hybrid systems \sep Piecewise-smooth dynamical systems

\maketitle

\section*{Nomenclature}
The mathematical symbols used throughout this paper are listed in Table \ref{tab 1}.

\begin{table}[H]
\centering
\caption{Description of mathematical symbols}\label{tab 1}
    \begin{tabular}{@{}lll@{}}
    \textbf{Variable name} & \textbf{Symbols} & \textbf{Dims.}\\
    \hline
    \textbf{Scalars} & & \\
    Instant of impact state $\mathbf{x}_i$ & $t_i$ & $\mathbb{R}^{1}$\\
    Discontinuity function & $H(\mathbf{x})$ & $\mathbb{R}^{1}$\\
    Flight time for $\mathbf{x}_0$ to reach $\Sigma_2$ at $\mathbf{x}_2$ & $\delta$ & $\mathbb{R}^{1}$\\
    First-order flight time for $\mathbf{x}_0$ to reach $\Sigma_2$ at $\mathbf{x}_2$ & $\delta_1$ & $\mathbb{R}^{1}$\\
    Positive root of the second-order flight time for $\mathbf{x}_0$ to reach $\Sigma_2$ at $\mathbf{x}_2$ & $\delta_+$ & $\mathbb{R}^{1}$\\
    Negative root of the second-order flight time for $\mathbf{x}_0$ to reach $\Sigma_2$ at $\mathbf{x}_2$ & $\delta_-$ & $\mathbb{R}^{1}$\\
    Components of perturbation state $\mathbf{y} \in \mathbb{R}^n$ & $y_i$ & $\mathbb{R}^{1}$\\
    Scaling factor of perturbation state $\mathbf{y}$ & $r_0$ & $\mathbb{R}^{1}$\\
    Quadratic function of flight time $\delta$ & $G$ & $\mathbb{R}^{1}$\\
    Coefficient of restitution & $r$ & $\mathbb{R}^{1}$\\
    Components of reset map $\mathbf{R}(\mathbf{x}_i)$ & $r_{1,2}$ & $\mathbb{R}^{1}$\\
    Position component of the rigid barrier & $\sigma$ & $\mathbb{R}^{1}$\\
    \hline
    \textbf{Vectors} & & \\
    Generalized state & $\mathbf{x}$ & $\mathbb{R}^{n}$\\
    Generalized velocity vector field & $\mathbf{F}(\mathbf{x})$ & $\mathbb{R}^{n}$\\
    Perturbation to $\mathbf{x}$ & $\mathbf{y}$ & $\mathbb{R}^{n}$\\
    Initial state $\mathbf{x}$ post transients on periodic orbit & $\mathbf{x}_p$ & $\mathbb{R}^{n}$\\
    Perturbation state to $\mathbf{x}_p$ & $\hat{\mathbf{x}}$ & $\mathbb{R}^{n}$\\
    State $\mathbf{x}$ at impact & $\mathbf{x}_i$ & $\mathbb{R}^{n}$\\
    Reset map & $\mathbf{R}(\mathbf{x}_i)$ & $\mathbb{R}^{n}$\\
    Perturbed state at impact & $\mathbf{x}_0$ & $\mathbb{R}^{n}$\\
    Perturbed state on $\Sigma_2$ & $\mathbf{x}_2$ & $\mathbb{R}^{n}$\\
    Reset map of perturbed state $\mathbf{R}(\mathbf{x}_2) \in \Sigma_2$ & $\mathbf{x}_3$ & $\mathbb{R}^{n}$\\
    TDM of perturbed state & $\mathbf{x}_4$ & $\mathbb{R}^{n}$\\
    Perturbation state during impact & $\mathbf{y}_-$ & $\mathbb{R}^{n}$\\
    TDM of perturbation state & $\mathbf{y}_+$ & $\mathbb{R}^{n}$\\
    \hline
    \textbf{Matrices} & & \\
    Jacobian matrix of $\mathbf{F}$ & $\nabla\mathbf{F}(\mathbf{x})^T$ & $\mathbb{R}^{n \times n}$\\
    Hessian matrix of $i^{th}$ component of $\mathbf{F}$ & $H_i$ & $\mathbb{R}^{n \times n}$\\
    Hessian matrix of $i^{th}$ component of $\mathbf{R}$ & $\tilde{H}_i$ & $\mathbb{R}^{n \times n}$\\
    First-order saltation matrix & $\mathbf{S}$ & $\mathbb{R}^{n \times n}$\\
    Higher-order saltation matrix & $\mathbf{S}_2$ & $\mathbb{R}^{n \times n}$\\
    Higher-order saltation matrix after $i^{th}$ impact &$\mathbf{S}^i_2$ & $\mathbb{R}^{n \times n}$\\
    Perturbation matrix at impact & $\mathbf{Y}_{-,\text{impact}}$ & $\mathbb{R}^{n \times n}$\\
    Perturbation matrix after impact & $\mathbf{Y}_{+,\text{impact}}$ & $\mathbb{R}^{n \times n}$\\
    State transition matrix before $i^{th}$ impact & $\mathbf{\Phi}_i$ & $\mathbb{R}^{n \times n}$\\
    Global state transition or monodromy matrix & $\mathbf{\Phi}$ & $\mathbb{R}^{n \times n}$\\
    \hline
    \textbf{Sets} & & \\
    Initial Poincar\'e section on periodic orbit & $\Sigma_1$ & $\mathbb{R}^{n - 1}$\\
    Discontinuity boundary & $\Sigma_2$ & $\mathbb{R}^{n - 1}$
    \end{tabular}
\end{table}

\section{Introduction}
\noindent Dynamical systems are often subjected to mechanical impacts, leading to possible compromise in performance and safety. 
These impacts arise due to wear and tear with limited clearance between components or loosened joints in a wide range of systems, including gear assemblies \cite{kahraman1990non,karagiannis1991theoretical}, impact print hammers \cite{hendriks1983bounce,tung1988method}, walking robots \cite{holmes2006dynamics} and heat exchanger tubes\cite{goyder1989study,paidoussis1992cross}. Similar effects occur in submerged and floating buoyant vessels impacting rigid harbours during ship grounding \cite{virgin2009some,ibrahim2014recent,xue2023nonlinear}, metal cutting \cite{wiercigroch2001sources}, drilling and milling \cite{krivtsov1999dry,wiercigroch1999material} and many-body interaction systems involving friction \cite{feeny1992nonsmooth,fan2020discontinuous}. The dynamics of these systems are often analyzed using the canonical form of impact oscillators \cite{shaw1983periodically,shaw1983periodicallythesis,bishop1994impact,jiang2016geometrical,luo2005periodic,chillingworth2013periodic,whiston1987vibro,whiston1987global,foale1992dynamical,foale1994bifurcations,ma2006border}, mathematically modelled as piecewise-smooth (PWS) dynamical systems \cite{bernardo2008piecewise,awrejcewicz2003bifurcation,belykh2023beyond,brogliato1999nonsmooth}. A wide range of theoretical \cite{nordmark1991non,nordmark1992effects,nusse1992border,simpson2020nordmark,nordmark1997universal,ma2008nature} and experimental \cite{wiercigroch1998experimental,banerjee2009invisible,ing2006dynamics,ing2008experimental,ing2010bifurcation,pavlovskaia2010complex} investigations of impact oscillators reveal occurrences of many dynamically rich phenomena, collectively known as discontinuity-induced bifurcations (DIBs) \cite{bernardo2008piecewise}. Some examples include period adding cascades \cite{piiroinen2004chaos,oestreich1996bifurcation,oestreich1997analytical,popp1999numerical,shaw1983periodically,fredriksson2000normal,rounak2020bifurcations}, grazing bifurcations \cite{banerjee2009invisible,jiang2017grazing,chin1994grazing}, sliding motion \cite{di2002bifurcations}, chattering sequence \cite{budd1994chattering}, narrow band and robust chaos \cite{banerjee2009invisible,thompson1983chaotic,hassouneh2004robust} and coexisting attractors \cite{chawla2024wake}. Unstable, chaotic vibrations resulting from DIBs can be undesirable or harmful for safety or performance, making their analysis essential for the design, performance and control.

DIBs primarily occur when an orbit grazes the discontinuity boundary \cite{bernardo2008piecewise}. Nordmark \cite{nordmark1991non} showed that in such conditions, orbits near the discontinuity boundary stretch in the phase space, resulting in large amplitude oscillations known as square-root singularity.  Away from this grazing condition, the evolution of perturbations near the discontinuity boundary is determined by a transverse discontinuity mapping (TDM) \cite{bernardo2008piecewise}. Fredriksson and Nordmark \cite{fredriksson2000normal} had derived the normal form of the TDM using first-order approximations. This TDM depends on the time taken for orbits to reach the discontinuity boundary. For an $n$ dimensional hybrid dynamical system $\mathbf{x} \in \mathbb{R}^{n}$ obeying $\dot{\mathbf{x}} = \mathbf{F}(\mathbf{x}, \mathbf{u}, t)$, a first-order Taylor series expansion \cite{fredriksson2000normal,leine2000bifurcations} approximates the time difference $\delta_1$ between impacts as
\begin{equation} \label{eq 1}
    \delta_1 = -\frac{\nabla H(\mathbf{x}_i)^T \cdot \mathbf{y}_-}{\nabla H(\mathbf{x}_i)^T \cdot \mathbf{F}(\mathbf{x}_i)}
\end{equation}
where $\mathbf{F}(\mathbf{x}, \mathbf{u}, t)$ is the governing vector field, $\mathbf{u}$ are parameters of the system and overdot represents derivative with respect to time $t$. The discontinuity boundary $\mathbf{\Sigma_2}$ is modelled by a scalar function $H(\mathbf{x})$ satisfying $\Sigma_2 = \{\mathbf{x}_i \in \mathbb{R}^n : H(\mathbf{x}_i) = 0\}$ and $\mathbf{y}_-$ is the perturbation vector during an impact of the primary state at $\mathbf{x}(t_i) = \mathbf{x}_i$; $t_i$ being the instant of impact.

This first-order flight time $\delta_1$ has an inherent problem. The flight time $\delta_1$ in Eq. \eqref{eq 1} is a real-valued fraction that assumes all perturbations in the local neighbourhood of an impact state $\mathbf{x}_i$, reaches the discontinuity bondary and gets mapped to $\mathbf{y}_+ = \mathbf{S} \cdot \mathbf{y}_-$ while the primary state is mapped to $\mathbf{x}_i \rightarrow \mathbf{R}(\mathbf{x}_i)$ via the impact map $\mathbf{R}(\mathbf{x})$. Here, $\mathbf{S}$, known as the saltation matrix, is the first-order TDM \cite{fredriksson2000normal,leine2000bifurcations}. The saltation matrix is essentially a state-transition matrix between perturbation vectors before and after impact, \textit{i.e.}, $\mathbf{y}_-$ and $\mathbf{y}_+$. Additionally, another drawback of the first-order approximation of the flight time and the saltation matrix is that for low-velocity incidence and grazing orbits, the TDM gets stretched since the denominator diverges as $\nabla H(\mathbf{x}_i)^T \cdot \mathbf{F}(\mathbf{x}_i) = 0$. This also results in overestimating the mapped state away from the discontinuity boundary. 

This paper first demonstrates how impacts are incorrectly predicted for certain perturbations in the local neighbourhood of the discontinuity boundary. Specifically, the first-order approximation of the flight time $\delta_1$ and the saltation matrix $\mathbf{S}$ fails to account for cases where the perturbed orbit misses the discontinuity boundary. Next, this work addresses this shortcoming by considering a higher-order Taylor series approximation to derive the flight time $\delta$ and TDM. This results in a quadratic equation in $\delta$, implying that perturbations reach the discontinuity boundary only when the discriminant of the quadratic equation is non-negative, and is given by the real positive root $\delta_+$. The positive root of the derived higher-order flight time estimate comprises additional terms in the denominator that ensure that the TDM does not diverge for low-velocity impacts close to the grazing condition. The correction terms, not present in the first-order saltation matrix, ensure that the mapped state post-impact is not overestimated for low-velocity impacts and reaches the discontinuity boundary correctly. These limitations have been overlooked mainly because the existing literature focuses on discontinuity mappings near grazing like the zero-time (ZDM) \cite{yin2018higher,kundu2012singularities} and Poincar\'e section discontinuity map (PDM) \cite{di2001normal,di2002bifurcations,di2001grazing,simpson2020nordmark}. Additionally, the current literature \cite{bernardo2008piecewise} on higher-order approximations of the TDM relies on the first-order flight time $\delta_1$. A singularity is unavoidable in the first-order flight time due to the denominator of Eq. \eqref{eq 1} for low-velocity impacts. Discrepancies between dynamical behaviour derived from mapping techniques and direct numerical simulations become evident in complex, chaotic dynamics, degenerate bifurcations, or higher co-dimensional bifurcations \cite{yin2018higher}. This article resolves the discrepancy by deriving the TDM by combining a higher-order quadratic form of $\delta_+$ and a higher-order Taylor series approximation of the perturbed state in the local neighbourhood of $\mathbf{x}_i$.

Next, a higher-order TDM derived in this article poses a new challenge. Since the mapped state $\mathbf{y}_+$ is proportional to terms like $\delta^2_+$, $\mathbf{y}_-\cdot\mathbf{y}_-$, and $\delta_+ \mathbf{y}_-$, the TDM cannot be expressed as a state transition or saltation matrix like $\mathbf{S} \cdot \mathbf{y}_-$. Hence, a higher-order saltation matrix comprising quadratic terms cannot be expressed in a closed analytical form, and estimation of monodromy matrices and Floquet multipliers for stability analysis becomes difficult. This work addresses the problem by introducing a method to numerically obtain the saltation matrix comprising higher-order correction terms compatible with the first-order variational equation.

Incorrect estimation of impact occurrences using the first-order TDM has significant consequences for stability analysis based on Floquet multipliers \cite{floquet1883linear,nayfeh2008applied} and Lyapunov exponents (LEs) \cite{oseledec1968multiplicative,pesin1977characteristic,benettin1980Lyapunov1,bennetin1980Lyapunov2}. The monodromy matrix comprises products of state transition and saltation matrices multiplied in their order of occurrence. Any incorrect prediction of the occurrence of an impact alters the monodromy matrix, resulting in an inaccurate representation of the perturbation dynamics near the discontinuity boundary. Consequently, using a first-order linearization instead of the proposed higher-order approach may result in misleading conclusions regarding orbital stability. Moreover, when the time between two consecutive impacts is large, the separation between nearby orbits can increase, potentially causing the perturbation vector to miss the discontinuity boundary while the primary orbit undergoes an impact. In such cases, the monodromy matrix should remain unaffected, as no saltation matrix should be applied. However, the first-order TDM, which assumes a real-valued flight time, cannot capture this scenario. In contrast, a higher-order TDM and saltation matrix correctly predicts when impacts occur. This ensures an accurate monodromy matrix and its corresponding Floquet multipliers.

Similarly, the computation of LEs in hybrid dynamical systems depends on the first-order TDM and saltation matrix \cite{lamba1994scaling,muller1995calculation,de2004calculation,jin2006method,leine2012non,mandal2013automated,stefanski2000estimation,stefanski2005evaluation,dabrowski2012estimation,li2018chaotic,balcerzak2020determining}. In impact oscillators, LEs are typically calculated after multiple impacts to eliminate transient effects. Therefore, an incorrect prediction of an impact of the perturbation vector accumulates over successive applications of the first-order saltation matrix, resulting in less accurate perturbation dynamics. Instead, the higher-order TDM provides better estimates of the true behaviour of perturbed orbits near the discontinuity boundary.

This paper thus presents a generalized method to calculate the flight times for impact of orbits in the local neighbourhood of the discontinuity boundary and the corresponding TDM by incorporating higher-order corrections terms up to $\mathcal{O}(\delta_+^2, \delta_+ \mathbf{y}_-, \mathbf{y}_- \cdot \mathbf{y}_-)$. The derivations can be extended to accommodate correction terms greater than $\mathcal{O}(2)$ if necessary, following the approach presented here, and subsequently applied to autonomous and non-autonomous hybrid systems with multiple barriers. Next, this article proposes a method to obtain higher-order saltation matrices numerically. The higher-order TDMs are validated by conducting a stability analysis of impact oscillators with multiple barriers using Floquet multipliers and LEs.

The article is structured as follows. In Sec. \ref{sec 2}, the higher-order TDM for a hybrid dynamical system is derived. Section \ref{sec 3} compares the higher-order results with the first-order linearized approaches for two representative impact oscillators with multiple impacts. In Sec. \ref{sec 4}, the higher-order TDM is implemented to obtain the Lyapunov spectrum. Section \ref{sec 5} presents the methodology to numerically evaluate a higher-order saltation matrix. Section \ref{sec 6} presents a method to construct a monodromy matrix using the higher-order saltation matrix, followed by an eigenvalue analysis of the respective impact oscillators. Section \ref{sec 7} investigates the stability analysis of the impact oscillator for which period-adding cascades are observed. Section \ref{sec 8} summarises the principal outcomes of this study.

\section{Mathematical formulation} \label{sec 2}

\begin{figure}[!htb]
    \centering
    \includegraphics[scale = 1.5]{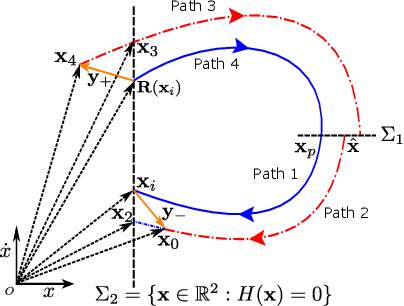}
    \caption{A schematic of phase portraits of two nearby trajectories exhibiting impact at $\Sigma_{2}$. The blue line shows the actual trajectory. The red dashed line denotes the perturbed trajectory. The horizontal black dashed line $\Sigma_1$ denotes the section from which both the trajectories are initiated, and $\Sigma_2$ denotes the discontinuity boundary. The amber line denotes the perturbation vector.}
    \label{fig 1}
\end{figure}

\noindent The generalized form of a dynamical system represented by the state $\mathbf{x}(t)$, where $ \mathbf{x} \in \mathbb{R}^n$, and its corresponding variational form, post transients, can be expressed in the state space form as
\begin{align} \label{eq 2}
    \dfrac{d\mathbf{x}}{dt} &= \mathbf{F}(\mathbf{x}), \\
    \dfrac{d\mathbf{y}}{dt} &\approx \mathbf{\nabla} \mathbf{F}(\mathbf{x})^T\cdot\mathbf{y}  + \mathcal{O}(||\mathbf{y}||^2)\nonumber
\end{align}
where $\mathbf{y}$ is a perturbation to the state $\mathbf{x}$. The variational form governs the dynamics of orbits in the local neighbourhood of $\mathbf{x}$ and is derived using first-order Taylor approximations. The state space form is also valid for a non-autonomous system with explicit time dependence, \textit{i.e.}, $\mathbf{F}(\mathbf{x}, t)$. The additional phase variable, $t$, can be included in $\mathbf{x}$ such that $\dot{t} = 1$, making the new state $\mathbf{x}^{\prime}$ $n+1$ dimensional. Therefore, the derivations presented in this section apply to both autonomous and non-autonomous systems. In general, for PWS systems, nearby perturbed trajectories reach the discontinuity boundary at different instants of time (see Fig. \ref{fig 1}). Implementing a linearized TDM to estimate flight times can result in overestimating the mapped perturbed state near the discontinuity boundary. Higher-order terms must be considered to accurately map perturbed trajectories for low-velocity impacts, as demonstrated below.

The evolution of two nearby trajectories is depicted in the simplistic case of a 2-dimensional state space (Fig. \ref{fig 1}). Two closely spaced states $\mathbf{x}_p$ and $\mathbf{\hat{x}}$ are initiated together from the Poincar\'e section $\Sigma_{1}$. Here, $\mathbf{\hat{x}}$ represents a perturbed trajectory from $\mathbf{x}_p$, \textit{i.e.} $\mathbf{\hat{x}} = \mathbf{x}_p + \mathbf{y}$, where $\mathbf{y}$ is an infinitesimal perturbation. After evolving in time, at $t = t_i$, the orbit $\mathbf{x}_p$ impacts the rigid surface represented by the discontinuity boundary  $\Sigma_2$ at $\mathbf{x}_i$. At this instant, when the orbit, initiated from $\mathbf{x}_p$ impacts the surface $\Sigma_{2} = \{ \mathbf{x}_i \in \mathbb{R}^n: H(\mathbf{x}_i) = 0\}$ , the trajectory gets mapped to $\mathbf{R}(\mathbf{x}_i)$. Here, $\mathbf{R}(\mathbf{x})$ is an impact map based on a restitutive law, whereas $H(\mathbf{x}) = 0$ models the impacting condition. The impact or reset map $\mathbf{R}(\mathbf{x})$ is a typical example of a non-smooth event that arises in PWS dynamical systems. Applying the impact map $\mathbf{R}(\mathbf{x}_0)$ to the perturbed trajectory $\mathbf{\hat{x}}(t_i) = \mathbf{x}_0$ at the instant of impact $t_i$ will result in an incorrect prediction of state since $\mathbf{x}_0$ has not yet reached the impacting surface $\Sigma_{2}$ at $t_i$. Thus, the difference in the flight times of the two paths $\mathbf{x}_i$ and $\mathbf{x}_0$ needs to be considered. The corresponding derivation of the flight time with higher-order approximations, is obtained next. Using the flight time difference, a higher-order TDM is derived, that accurately maps the state $\mathbf{x}_0$ to $\mathbf{x}_4$. It is important to note that $\mathbf{x}_4$ now lies in the forbidden region $H(\mathbf{x})<0$ but eventually evolves in time to reach the barrier at $H(\mathbf{x}_3) = 0$. Post impact, the perturbed trajectory when initiated at $\mathbf{x}_4$, would take the same flight time to arrive at $\mathbf{x}_3$ on $\Sigma_2$ as the perturbed trajectory at $\mathbf{x}_0$ would take to reach $\mathbf{x}_2$, followed by the mapping $\mathbf{R}(\mathbf{x}_2)$ on $\Sigma_2$ to reach $\mathbf{x}_3$. This ensures that local trajectories obey a zero-time mapping to the discontinuity boundary, and the time difference is conserved. The mapping of the perturbed path from $\mathbf{x}_0$ to $\mathbf{x}_4$ at the instant of the impact of the unperturbed trajectory is the required TDM; see Fig. \ref{fig 1}. The saltation matrix is essentially a first-order approximation of the state transition from $\mathbf{x}_0$ to $\mathbf{x}_4$. A higher-order approximation of the flight time and the TDM is presented next. 

Let $t_i$ represent the instant of the impact of the unperturbed trajectory. A Taylor expansion of $\mathbf{x}(t)$ about $t = t_i$ and $\mathbf{F}(\mathbf{x})$ about $\mathbf{x} = \mathbf{x}_0$ up to $\mathcal{O}(2)$ gives Eqs. \ref{eq 3} and \eqref{eq 4}.
\begin{align} \label{eq 3}
    \mathbf{x}(t) = \mathbf{x}(t_i) + \Delta t \ \mathbf{F}(\mathbf{x}_0) + \dfrac{1}{2}\Delta t^2 \  \mathbf{\nabla}\mathbf{F}(\mathbf{x}(t_i))^T\cdot\mathbf{F}(\mathbf{x}(t_i)) + \mathcal{O}(3)
\end{align}
\begin{align} \label{eq 4}
    \mathbf{F}(\mathbf{x}) = \mathbf{F}(\mathbf{x}_0) + \mathbf{\nabla}\mathbf{F}(\mathbf{x}_0)^T\cdot\mathbf{\Delta x} + \dfrac{1}{2}\begin{bmatrix} \mathbf{\Delta x}^T\cdot H_1\cdot\mathbf{\Delta x}\\ \mathbf{\Delta x}^T\cdot H_2\cdot\mathbf{\Delta x}\\ \end{bmatrix} + \mathcal{O}(3)
\end{align}
Here, $H_1$ and $H_2$ denote the Hessian matrices of each component $f_i(\mathbf{x})$ of $\mathbf{F}(\mathbf{x})$ {\it i.e.,} $H_i$ defined as the Jacobian of $\mathbf{\nabla} f_i (\mathbf{x})$ or $\mathbf{\nabla}(\mathbf{\nabla}f_i(\mathbf{x}))^T$. Let $\delta$ be the time taken for the perturbed trajectory $\mathbf{\hat x}$ to reach impact surface $\Sigma_{2}$ from $\mathbf{x}(t_i) = \mathbf{x}_0$ to $\mathbf{x}(t_i + \delta) = \mathbf{x}_2$. Therefore, $\mathbf{x}_2$ is approximated by taking a Taylor expansion of $\mathbf{\hat x}$ along path 2 about $t = t_i$ (\textit{i.e.,} $\mathbf{x}(t_i) = \mathbf{x}_0)$ evaluated at $t = t_i + \delta$ up to $\mathcal{O}(2)$ giving
\begin{align} \label{eq 5}
    \mathbf{x}_2 = \mathbf{x}_0 + \delta \ \mathbf{F}(\mathbf{x}_0) + \dfrac{1}{2} \delta^2 \ \mathbf{\nabla} \mathbf{F}(\mathbf{x}_0)^T\cdot\mathbf{F}(\mathbf{x}_0) + \mathcal{O}(3)
\end{align}
$\mathbf{F}(\mathbf{x}_0)$ in Eq. \eqref{eq 5} is approximated by a Taylor expansion of $\mathbf{F}(\mathbf{x})$ along path 1 about $\mathbf{x}_i$ and evaluated at $\mathbf{x} = \mathbf{x}_0$. Retaining terms up to $\mathcal{O}(2)$, $\mathbf{x}_2$ becomes Eq. (\ref{eq 6})
\begin{align} \label{eq 6}
    \mathbf{x}_2 = \mathbf{x}_0 + \delta \ [\mathbf{F}(\mathbf{x}_i) + \mathbf{\nabla} \mathbf{F}(\mathbf{x}_i)^T\cdot\mathbf{y}_{-}] + \dfrac{1}{2} \delta^2 \ \mathbf{\nabla} \mathbf{F}(\mathbf{x}_i)^T\cdot\mathbf{F}(\mathbf{x}_i) + \mathcal{O}(3)
\end{align}

\noindent where $\mathbf{y}_-$ is the perturbation vector at the instant of impact (\textit{i.e.} $\mathbf{x}_0 = \mathbf{x}_i + \mathbf{y}_-$). 

\begin{theorem}
For all $\mathbf{x}_i \in \Sigma_2$ at $t_i$, the flight time $\delta$ taken by $\mathbf{y}_-$ to reach $\mathbf{x}_2 \in \Sigma_2$ satisfies the quadratic equation, $G(\delta, \mathbf{x}_i, \mathbf{y}_-) = A\delta^2 + B \delta + C = 0$ where the scalars $A$, $B$, $C$ are defined as
\begin{align} \label{eq 7}
    A &= \mathbf{\nabla}H(\mathbf{x}_i)^T\cdot\mathbf{\nabla}\mathbf{F}(\mathbf{x}_i)^T\cdot\mathbf{F}(\mathbf{x}) + \mathbf{F}(\mathbf{x}_i)^T\cdot\mathbf{\nabla}(\mathbf{\nabla}H(\mathbf{x}_i))^T\cdot\mathbf{F}(\mathbf{x}_i), \\
    B &= 2\mathbf{\nabla} H(\mathbf{x}_i)^T\cdot\mathbf{F}(\mathbf{x}_i) + 2\mathbf{\nabla}H(\mathbf{x}_i)^T\cdot\mathbf{\nabla}\mathbf{F}(\mathbf{x}_i)^T\cdot\mathbf{y} + \mathbf{y}^T\cdot\mathbf{\nabla}(\mathbf{\nabla}H(\mathbf{x}_i))^T\cdot\mathbf{F}(\mathbf{x}_i) + \mathbf{F}(\mathbf{x}_i)^T\cdot\mathbf{\nabla}(\mathbf{\nabla}H(\mathbf{x}_i))^T\cdot\mathbf{y} \nonumber, \\
    C &= \mathbf{y}^T\cdot\mathbf{\nabla}(\mathbf{\nabla}H(\mathbf{x}_i))^T\cdot\mathbf{y} + 2 \mathbf{\nabla}H(\mathbf{x}_i)^T\cdot \mathbf{y} \nonumber .
\end{align}
\end{theorem}

\begin{proof}
The equation for the impacting surface $H(\mathbf{x})$ expanded about the state at impact $\mathbf{x}_i$ up to $\mathcal{O}(2)$ and evaluated at $\mathbf{x} = \mathbf{x}_2$ is given by
\begin{align} \label{eq 8}
    H(\mathbf{x}_2) = \ H(\mathbf{x}_i) + \mathbf{\nabla} H(\mathbf{x}_i)^T\cdot(\mathbf{x}_2 - \mathbf{x}_i) + \dfrac{1}{2}(\mathbf{x}_2 - \mathbf{x}_i)^T\cdot\mathbf{\nabla}(\mathbf{\nabla}H)^T\cdot(\mathbf{x}_2 - \mathbf{x}_i) + \mathcal{O}(3)
\end{align}

\noindent Substituting Eq. \eqref{eq 6} in Eq. \eqref{eq 8} and equating $H(\mathbf{x}_i) = H(\mathbf{x}_2) = 0$ (since $\mathbf{x}_i$ and $\mathbf{x}_2$ lie on the impacting surface $\Sigma_{2}$), the time difference in impact between two closely spaced trajectories, \textit{i.e.} $\delta$ can be solved up to $\mathcal{O}(2)$. This results in a quadratic equation in $\delta$ given by
\begin{align}\label{eq 9}                           \delta^2\Big(&\mathbf{\nabla}H(\mathbf{x}_i)^T\cdot\mathbf{\nabla}\mathbf{F}(\mathbf{x}_i)^T\cdot\mathbf{F}(\mathbf{x}) + \mathbf{F}(\mathbf{x}_i)^T\cdot\mathbf{\nabla}(\mathbf{\nabla}H(\mathbf{x}_i))^T\cdot\mathbf{F}(\mathbf{x}_i)\Big) + \delta \Big( 2\mathbf{\nabla} H(\mathbf{x}_i)^T\cdot\mathbf{F}(\mathbf{x}_i) \\&+ 2\mathbf{\nabla}H(\mathbf{x}_i)^T\cdot\mathbf{\nabla}\mathbf{F}(\mathbf{x}_i)^T\cdot\mathbf{y} + \mathbf{y}^T\cdot\mathbf{\nabla}(\mathbf{\nabla}H(\mathbf{x}_i))^T\cdot\mathbf{F}(\mathbf{x}_i) + \mathbf{F}(\mathbf{x}_i)^T\cdot\mathbf{\nabla}(\mathbf{\nabla}H(\mathbf{x}_i))^T\cdot\mathbf{y} \Big) \nonumber \\&\quad+ \mathbf{y}^T\cdot\mathbf{\nabla}(\mathbf{\nabla}H(\mathbf{x}_i))^T\cdot\mathbf{y} + 2 \mathbf{\nabla}H(\mathbf{x}_i)^T\cdot \mathbf{y} + \mathcal{O}(3) = G(\delta, \mathbf{x}_i, \mathbf{y}_-) = 0 \nonumber
\end{align}
\end{proof}

%Since, a higher-order approximation of $\delta$ results in a quadratic equation, Eq. \eqref{eq 8} can be written as $G(\delta, \mathbf{x}_i, \mathbf{y}_-) = A\delta^2 + B \delta + C = 0$ where the scalars $A$, $B$, $C$ are defined as
%\begin{align} \label{eq 9}
%    A &= \mathbf{\nabla}H(\mathbf{x}_i)^T\cdot\mathbf{\nabla}\mathbf{F}(\mathbf{x}_i)^T\cdot\mathbf{F}(\mathbf{x}) + \mathbf{F}(\mathbf{x}_i)^T\cdot\mathbf{\nabla}(\mathbf{\nabla}H(\mathbf{x}_i))^T\cdot\mathbf{F}(\mathbf{x}_i) \\
%    B &= 2\mathbf{\nabla} H(\mathbf{x}_i)^T\cdot\mathbf{F}(\mathbf{x}_i) + 2\mathbf{\nabla}H(\mathbf{x}_i)^T\cdot\mathbf{\nabla}\mathbf{F}(\mathbf{x}_i)^T\cdot\mathbf{y} + \mathbf{y}^T\cdot\mathbf{\nabla}(\mathbf{\nabla}H(\mathbf{x}_i))^T\cdot\mathbf{F}(\mathbf{x}_i) + \mathbf{F}(\mathbf{x}_i)^T\cdot\mathbf{\nabla}(\mathbf{\nabla}H(\mathbf{x}_i))^T\cdot\mathbf{y} \nonumber \\
%    C &= \mathbf{y}^T\cdot\mathbf{\nabla}(\mathbf{\nabla}H(\mathbf{x}_i))^T\cdot\mathbf{y} + 2 \mathbf{\nabla}H(\mathbf{x}_i)^T\cdot \mathbf{y} \nonumber
%\end{align}

\begin{lemma}\label{lemma 2}
    The perturbed state $\mathbf{x}_i + \mathbf{y_-}$ reaches the discontinuity boundary $\Sigma_2 = \{\mathbf{x} \in \mathbb{R}^n:H(\mathbf{x}) = 0\}$, i.e., there exists a $\mathbf{x}_2 \in \Sigma_2$, after the flight time $\delta$ iff the quadratic equation $G(\delta, \mathbf{x}_i, \mathbf{y}_-) = A\delta^2 + B \delta + C = 0$ has a real solution. This is ensured by the discriminant condition, i.e., $B^2 \geq 4AC$. In this case, the discriminant is non-negative, ensuring $\delta \in \mathbb{R}$; otherwise, $\delta$ is imaginary and no impact occurs.
\end{lemma}

\begin{lemma}\label{lemma 3}
When $B^2 \geq 4AC$, the flight time is given by the positive root of $G(\delta, \mathbf{x}_i, \mathbf{y}_-) = 0$, \textit{i.e.},
\begin{equation} \label{eq 10}
    \delta_+ = -\frac{B}{2A}\Bigg(1 - \sqrt{1 - \frac{4AC}{B^2}}\Bigg).
\end{equation}
\end{lemma}

\begin{proof}
The roots of $G(\delta, \mathbf{x}_i, \mathbf{y}_-) = 0$ are given by the solution $-\dfrac{B}{2A}\bigg(1 \pm \sqrt{1 - \dfrac{4AC}{B^2}}\bigg)$ that ensures that $H(\mathbf{x}_2) = 0$ and implies that perturbed trajectories reach the discontinuity boundary after time $t_i + \delta$. Only the positive root of $\delta$ captures the limiting case when $\mathbf{x}_2 \rightarrow \mathbf{x}_i$ when $\mathbf{y}_- \rightarrow 0$ and, hence, $\delta_+ \rightarrow 0$.
\end{proof}

A comparison of Eq. \eqref{eq 9} with Eq. \eqref{eq 1} highlights two essential features. First, imaginary roots of Eq. \eqref{eq 9} can exist and would imply that the perturbed orbits do not reach the discontinuity boundary. This contradicts the results of the first-order approximation of $\delta$. Second, the magnitude of the perturbation vector $\mathbf{y}_-$ should lie within a critical range, which has been described in the later paragraphs, for impacts to occur. This critical range is obtained from the condition $B^2 \geq 4AC$ that ensures that the discriminant of Eq. \eqref{eq 10} is non-negative. Therefore, the correct flight time is given by the positive root of $G(\delta, \mathbf{x}_i, \mathbf{y}_-) = 0$. Geometrically, the discriminant condition suggests that when perturbation vectors become too large, they do not reach the discontinuity boundary and continue evolving in the region where $H(\mathbf{x}) > 0$. On the contrary, the first-order approximation in $\delta = \delta_1$ predicts that impacts occur for all perturbations. Therefore, this phenomenon cannot be captured while implementing a linearized approximation of the TDM.

\begin{theorem}\label{theorem 4}
At the impact time $t_i$, the perturbed state $\mathbf{x}_0$ is mapped to the post-impact state $\mathbf{x}_4$ given by the higher-order TDM,
\begin{align} \label{eq 11}
    \mathbf{x}_4 = \mathbf{R}(\mathbf{x}_i) &+ \mathbf{\nabla}\mathbf{R}(\mathbf{x}_i)^T\cdot\mathbf{y}_- +\delta_+ \mathbf{\nabla}\mathbf{R}(\mathbf{x}_i)^T\cdot\mathbf{F}(\mathbf{x}_i) - \delta_+ \mathbf{F}(\mathbf{R}(\mathbf{x}_i)) + \delta_+ \mathbf{\nabla}\mathbf{R}(\mathbf{x}_i)^T\cdot\mathbf{\nabla}\mathbf{F}(\mathbf{x}_i)^T\cdot\mathbf{y}_- \\&+ \dfrac{1}{2} \delta_+^2 \mathbf{\nabla}\mathbf{R}(\mathbf{x}_i)^T\cdot\mathbf{\nabla}\mathbf{F}(\mathbf{x}_i)^T\cdot\mathbf{F}(\mathbf{x}_i) -\delta_+ \mathbf{\nabla}\mathbf{F}(\mathbf{R}(\mathbf{x}_i))^T\cdot\mathbf{\nabla}\mathbf{R}(\mathbf{x}_i)^T\cdot\mathbf{y}_- 
    \nonumber \\&\quad- \delta_+^2 \mathbf{\nabla}\mathbf{F}(\mathbf{R}(\mathbf{x}_i))^T\cdot\mathbf{\nabla}\mathbf{R}(\mathbf{x}_i)^T\cdot\mathbf{F}(\mathbf{x}_i) + \dfrac{1}{2} \delta_+^2 \mathbf{\nabla}\mathbf{F}(\mathbf{R}(\mathbf{x}_i))^T\cdot\mathbf{F}(\mathbf{R}(\mathbf{x}_i)) \nonumber \\ 
    &\qquad+ \dfrac{1}{2} \begin{bmatrix} [\mathbf{y}_- + \delta_+ \mathbf{F}(\mathbf{x}_i)]^T\cdot\mathbf{\nabla}(\mathbf{\nabla}r_1)^T\cdot[\mathbf{y}_- + \delta_+ \mathbf{F}(\mathbf{x}_i)] \\ [\mathbf{y}_- + \delta_+ \mathbf{F}(\mathbf{x}_i)]^T\cdot\mathbf{\nabla}(\mathbf{\nabla}r_2)^T\cdot[\mathbf{y}_- + \delta_+ \mathbf{F}(\mathbf{x}_i)] \end{bmatrix} + \mathcal{O}(3) \nonumber
\end{align}
\end{theorem}

\begin{proof}
Given that \textbf{Lemma \ref{lemma 2}} (discriminant condition) is satisfied, applying \textbf{Lemma \ref{lemma 3}} (positive root of flight time), the TDM maps $\mathbf{x}_0$ to the post-impact state $\mathbf{x}_4$ such that $\mathbf{x}_4(t_i + \delta_+) = \mathbf{x}_3$ where $\mathbf{x}_3 = \mathbf{R}(\mathbf{x}_2)$; see Fig. \ref{fig 1}. This ensures that any perturbed trajectory initiated from $\mathbf{\hat{x}}(t)$ is correctly mapped to $\mathbf{x}_3$ on the discontinuity boundary $\Sigma_{2}$ at time $t_i + \delta_+$. The closed-form of $\mathbf{x}_4$ is approximated by expanding $\mathbf{x}_3$ along path 3 about $t = 0$. In the absence of $\Sigma_2$, $\mathbf{x}_4$ would naturally evolve to $\mathbf{x}_3$ after time $\delta_+$. Thus $\mathbf{x}_4$ is obtained by expanding and evaluating $\mathbf{x}_3$ backwards in time $t = -\delta_+$ and becomes
\begin{align} \label{eq 12}
    \mathbf{x}_4 = \mathbf{x}_3 - \delta_+ \ \mathbf{F}(\mathbf{x}_3) &+ \dfrac{1}{2} \delta_+^2 \ \mathbf{\nabla}\mathbf{F}(\mathbf{x}_3)^T\cdot\mathbf{F}(\mathbf{x}_3) + \mathcal{O}(3)
\end{align}

\noindent where $\mathbf{x}_3 = \mathbf{R}(\mathbf{x}_2)$. Expanding $\mathbf{R}(\mathbf{x})$ about $\mathbf{x}_i$ gives
\begin{align} \label{eq 13}
    \mathbf{R}(\mathbf{x}) = \mathbf{R}(\mathbf{x}_i) &+ \mathbf{\nabla}\mathbf{R}(\mathbf{x}_i)^T\cdot\mathbf{\Delta x} + \dfrac{1}{2}\begin{bmatrix} \mathbf{\Delta x}^T\cdot\tilde{H}_1\cdot\mathbf{\Delta x}\\ \mathbf{\Delta x}^T\cdot\tilde{H}_2\cdot\mathbf{\Delta x}\\ \end{bmatrix} + \mathcal{O}(3)
\end{align}

\noindent where $\mathbf{\Delta x} = \mathbf{x} - \mathbf{x}_i$ and $\tilde{H}_i$ are the Hessian matrices of each component $r_i(\mathbf{x})$ of $\mathbf{R}(\mathbf{x})$. These Hessian matrices are defined as the Jacobian of $\mathbf{\nabla}r_i(\mathbf{x})$ or $\mathbf{\nabla}(\mathbf{\nabla}r_i(\mathbf{x}))^T$. An approximation of $\mathbf{R}(\mathbf{x}_2)$ using Eq. \eqref{eq 6} up to $\mathcal{O}(2)$ results in the following expression
\begin{align} \label{eq 14}
    \mathbf{R}(\mathbf{x}_2) = \ \mathbf{R}(\mathbf{x}_i) &+ \mathbf{\nabla}\mathbf{R}(\mathbf{x}_i)^T\cdot\Big(\mathbf{y}_- + \delta_+ \mathbf{F}(\mathbf{x}_i) + \delta_+ \mathbf{\nabla}\mathbf{F}(\mathbf{x}_i)^T\cdot\mathbf{y}_-
    + \dfrac{1}{2} \delta_+^2 \mathbf{\nabla}\mathbf{F}(\mathbf{x}_i)^T\cdot\mathbf{F}(\mathbf{x}_i)\Big) \nonumber \\
    &\quad+ \dfrac{1}{2} \begin{bmatrix} [\mathbf{y}_- + \delta_+ \mathbf{F}(\mathbf{x}_i)]^T\cdot\mathbf{\nabla}(\mathbf{\nabla}r_1)^T\cdot[\mathbf{y}_- + \delta_+ \mathbf{F}(\mathbf{x}_i)] \\ [\mathbf{y}_- + \delta_+ \mathbf{F}(\mathbf{x}_i)]^T\cdot\mathbf{\nabla}(\mathbf{\nabla}r_2)^T\cdot[\mathbf{y}_- + \delta_+ \mathbf{F}(\mathbf{x}_i)] \end{bmatrix} + \mathcal{O}(3)
\end{align}
where $r_1$ and $r_2$ are the components of the map $\mathbf{R}(\mathbf{x})$. Terms up to $\mathcal{O}(2)$ is taken in $\mathbf{\Delta x} = \mathbf{x}_2 - \mathbf{x}_i$. Next, $\mathbf{F}(\mathbf{x}_3)$ in Eq. \eqref{eq 12} can be evaluated by expanding $\mathbf{F}(\mathbf{x})$ along path 4 about $\mathbf{R}(\mathbf{x}_i)$. Taking terms up to $\mathcal{O}(1)$ in $\mathbf{F}(\mathbf{x})$ yields
\begin{align} \label{eq 15}
    \mathbf{F}(\mathbf{x}_3) = \mathbf{F}(\mathbf{R}(\mathbf{x}_i)) + \mathbf{\nabla}\mathbf{F}(\mathbf{R}(\mathbf{x}_i))^T\cdot\Big( \mathbf{\nabla}\mathbf{R}(\mathbf{x}_i)^T\cdot\mathbf{y}_- + \delta_+ \ \mathbf{\nabla}\mathbf{R}(\mathbf{x}_i)^T\cdot\mathbf{F}(\mathbf{x}_i) \Big) + \mathcal{O}(2)
\end{align}
Now, substituting the expressions for $\mathbf{R}(\mathbf{x}_2)$, $\mathbf{F}(\mathbf{x}_3)$ in Eq. \eqref{eq 12}, the higher-order TDM of the perturbed trajectory from $\mathbf{x}_0$ to $\mathbf{x}_4$ at the instant of impact up to $\mathcal{O}(2)$ can be analytically found and is given by \eqref{eq 11}.
\end{proof}

Defining $\mathbf{y}_{-}$ and $\mathbf{y}_{+}$ as the perturbation vector between path 1 and path 2 before and after impact, we have $\mathbf{x}_4 - \mathbf{R}(\mathbf{x}_i) = \mathbf{y}_{+}$ and $\mathbf{x}_0 - \mathbf{x}_i = \mathbf{y}_-$. Therefore, the proposed higher-order correction maps $\mathbf{x}_i$ to $\mathbf{R}(\mathbf{x}_i)$, $\mathbf{x}_0$ to $\mathbf{x}_4$ and $\mathbf{y}_-$ to $\mathbf{y}_+$.

To obtain the widely accepted $\mathcal{O}(1)$ saltation matrix, only the $1^{st}$ order terms in the Eq. \eqref{eq 9} can be retained. This simplification leads to the following first-order $\delta_1$
\begin{equation} \label{eq 16}
    \delta_1 = -\dfrac{\mathbf{\nabla}H(\mathbf{x}_i)^T\cdot\mathbf{y}_-}{\mathbf{\nabla}H(\mathbf{x}_i)^T\cdot\mathbf{F}(\mathbf{x}_i)}
\end{equation}
This system was analytically formulated in \cite{fredriksson2000normal}. A substitution of Eq. \eqref{eq 16} in Eq. \eqref{eq 11} and retention of terms up to $\mathcal{O}(1)$ yields
\begin{align} \label{eq 17}
    \mathbf{x}_4 = \mathbf{R}(\mathbf{x}_i) + \mathbf{\nabla}\mathbf{R}(\mathbf{x}_i)^T\cdot\mathbf{y}_- + \dfrac{\Big( \mathbf{F}(\mathbf{R}(\mathbf{x}_i)) - \mathbf{\nabla}\mathbf{R}(\mathbf{x}_i)^T\cdot\mathbf{F}(\mathbf{x}_i) \Big)}{\mathbf{\nabla}H(\mathbf{x}_i)^T\cdot\mathbf{F}(\mathbf{x}_i)} \mathbf{\nabla}H(\mathbf{x}_i)^T\cdot\mathbf{y}_-
\end{align}
One can define a state transition matrix (STM) $\mathbf{S}$ that governs the mapping of $\mathbf{y}_-$ to $\mathbf{y}_+$ given by
\begin{equation} \label{eq 18}
    \mathbf{y}_+ = \mathbf{S} \cdot \mathbf{y}_-
\end{equation}
On substituting the above expression in Eq. \eqref{eq 17} and the relation between $\mathbf{x}_4$ and $\mathbf{R}(\mathbf{x}_i)$, the STM, also known as the saltation matrix, becomes
\begin{align} \label{eq 19}
    \mathbf{S} = &\mathbf{\nabla}\mathbf{R}(\mathbf{x}_i)^T + 
    \dfrac{\Big( \mathbf{F}(\mathbf{R}(\mathbf{x}_i)) - \mathbf{\nabla}\mathbf{R}(\mathbf{x}_i)^T\cdot\mathbf{F}(\mathbf{x}_i) \Big)}{\mathbf{\nabla}H(\mathbf{x}_i)^T\cdot\mathbf{F}(\mathbf{x}_i)} \otimes\mathbf{\nabla}H(\mathbf{x}_i)^T
\end{align}
The following sections compare the first-order and higher-order flight times and TDMs for the representative case of an impact oscillator.  

\section{Piecewise-smooth hybrid systems} \label{sec 3}
The derived higher-order mapping in Eqs. \eqref{eq 9} and \eqref{eq 11} are implemented to study the behaviour of non-smooth limit cycles of vibro-impacting oscillators with a rigid barrier. Two representative impact oscillators comprising single and multiple barriers, respectively, are chosen to investigate the accuracy of the proposed higher-order theory. 
\subsection{Impact oscillator}
\begin{figure}[!htb]
    \centering
    \includegraphics[scale = 1.1]{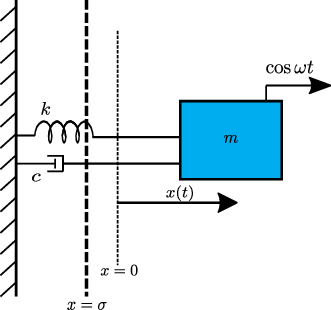}
    \caption{Periodically forced linear oscillator with barrier placed at $x = \sigma$.}
    \label{fig 2}
\end{figure}

\noindent Fig. \ref{fig 2} represents a classical harmonic oscillator with mass $m$, damping constant $c$ and stiffness $k$ subjected to an external harmonic forcing of frequency $\omega$. The corresponding non-dimensionalized governing equations for this impact oscillator \cite{de2004calculation} are
\begin{equation} \label{eq 20}
    \begin{cases}
    \ddot{x} + x + 2\xi \dot{x}= \cos{(\omega t)}, \quad \text{if} \ x > \sigma,\\
    \dot{x}(t_+) = -r\dot{x}(t_-) ,\quad \quad \quad \ \ \text{if} \ x(t_-) =  \sigma,
    \end{cases}
\end{equation}
where the undeformable impacting barrier is placed at $x(t_i) = \sigma$. At the instant of impact $t = t_i$, the oscillator undergoes an instantaneous reversal of velocity, which is modelled as $\dot{x}(t_{+}) = -r \, \dot{x}(t_{-}) = -r v$. Here, $t_{-}$ and $t_{+}$ are the instants before and after a collision, and $r \in (0, 1]$ depicts the coefficient of restitution.
The dynamics of perturbation vectors in the local neighbourhood of the primary state $\mathbf{x}$ is investigated next using Eqs. \eqref{eq 9} and \eqref{eq 11}. The first-order non-smooth discontinuity mapping is compared with the higher-order derived in Sec. \ref{sec 2}. Two closely spaced trajectories are initiated, post-transient effects, and these trajectories evolve according to the variational form away from the barrier. The integration is performed in Mathematica using its built-in ODE solver, NDSolve. An event detection routine is implemented for event detection \textit{i.e.,} when $x(t_-) = \sigma$. An accuracy of up to 15 digits after the decimal point is ensured to detect impacts with the barrier. 

Consider the evolution of orbits in the local neighbourhood of a steady state corresponding to $\xi = 2.0$, $\omega = 1.8$, and a barrier placed at $x = \sigma$ = -0.11. Two orbits $\mathbf{x}(t)$ and $\mathbf{x}(t) + \mathbf{y}(t)$ are initiated from $\mathbf{x}(3488.19) = [0.162579,0]$ and $\mathbf{y}(3488.19) = r_0/\sqrt{2}[1,1]$, where $r_0 = 0.007$ controls the magnitude of the normalized perturbation $1/\sqrt{2}[1, 1]$. The primary state impacts the barrier at $\mathbf{x}_i = [-0.11, -0.0577068]$ while perturbation vector reaches $\mathbf{y}_- = [0.00435243,-0.00115247]$ with norm $||\mathbf{y}_-|| = 0.00450243$. At this instant, the $\mathcal{O}(1)$ and $\mathcal{O}(2)$ approximations of flight time $\delta$, using Eqs \eqref{eq 16} and \eqref{eq 10} for the system Eq. \eqref{eq 20}, are given by, 
\begin{align}\label{eq 21}
    \delta^{\xi}_1 &= -\dfrac{y_1}{v}, \\
    \delta^{\xi}_{\pm} &= -\dfrac{v + y_2}{-\sigma - 2\xi v + \cos{(\omega t_i)}} \nonumber \\ 
        & \qquad \pm \dfrac{v + y_2}{-\sigma - 2\xi v + \cos{(\omega t_i)}} \sqrt{1-2\dfrac{(-\sigma - 2\xi v + \cos{(\omega t_i)})y_1}{(v + y_2)^2}} \nonumber 
\end{align}
Here, $\pm$ denotes the positive and negative roots of $G(\delta, \mathbf{x}_i, \mathbf{y}_-) = 0$, the impact state is $\mathbf{x}_i = [\sigma, v]$ at $t = t_i$ and $y_i$ are the components of the perturbation vector $\mathbf{y}$. The higher-order flight time approximation predicts impacts only occur for perturbations for which the discriminant of $\delta^{\xi}_+$ is positive. This is demonstrated next.

\begin{figure}[!]
    \centering
    \includegraphics[scale = 0.8]{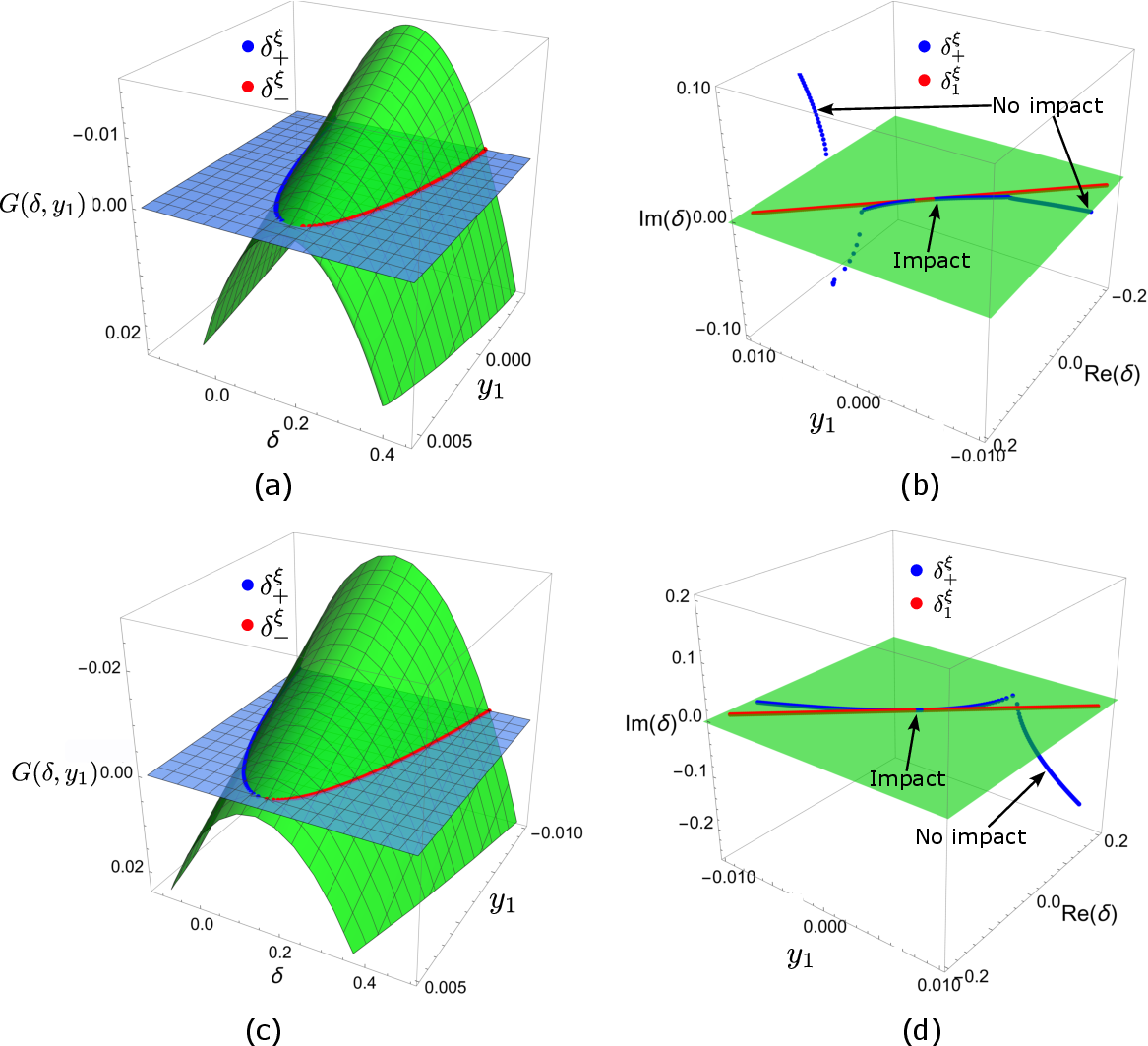}
    \caption{Figures depicting flight times of perturbations reaching the discontinuity barrier $\Sigma_2$. Perturbations only reach $\Sigma_2$ at the points where the two surfaces meet. The contour $G(\delta, y_1)$ is shown in green, and the blue plane represents $G(\delta, y_1) = 0$. The locus of the intersecting curve $\delta^{\xi}_{\pm}$ is the roots given in \eqref{eq 21}, and $\delta^{\xi}_{+}$ is the locus of points for which the impacts occur. Figures correspond to impact states $\mathbf{x}_i = [-0.11, -0.0577068]$ and $\mathbf{y}_- = [0.00435243, -0.00115247]$ at $t_i = 3489.83$. The green surface indicates $G(\delta, y_1)$ vs $\delta$ and $y_1$ with (a) norm of $\mathbf{y}_-$ fixed at $||\mathbf{y}_-|| = 0.00450243$ and (c) second component of $\mathbf{y}_-$ fixed at $y_2 = -0.00115247$. Imaginary and real parts of flight time $\delta$ vs $y_1$ with (b) norm of $\mathbf{y}_-$ fixed at $||\mathbf{y}_-|| = 0.00450243$ and (d) second component of $\mathbf{y}_-$ fixed at $y_2 = -0.00115247$. The impact barrier is placed at $\sigma = -0.11$ with system parameters $\xi = 2.0$ and $\omega = 1.8$.}
    \label{fig 3}
\end{figure}

Fig. \ref{fig 3} demonstrates, for a given impact occurrence at $\sigma$ where $\mathbf{x}_i$, $\mathbf{y}_-$ is known, the perturbation state $\mathbf{x}_i + \mathbf{y}_-$ can miss the discontinuity boundary depending on the magnitude of perturbation $||\mathbf{y}_-||$ and its components $y_1$, $y_2$. The results correspond to impact states $\mathbf{x}_i = [-0.11, -0.0577068]$, $\mathbf{y}_- = [0.00435243, -0.00115247]$ at $t_i=3489.83$ with $\xi = 2.0$ and $\omega = 1.8$. Fig. \ref{fig 3}(a) presents a contour plot, shown in green, of the surface $G(\delta, \mathbf{x}_i, \mathbf{y}_-)$ vs $\delta$ and $y_1$ where the first component of $\mathbf{y}_-$ is varied while keeping the norm fixed at $||\mathbf{y}_-|| = 0.00450243$. Impacts occur after a time elapse of $\delta$ for certain perturbed states $y_1$, and the solution is given by the curve formed from the intersection of the surface $G(\delta, \mathbf{x}_i, \mathbf{y}_-)$ (green surface) with the plane $G(\delta^{\xi}_{\pm}, \mathbf{x}_i, \mathbf{y}_-) = 0$ (blue plane). Eqs. \eqref{eq 21} gives the locus of these points on the curve, and they are shown as blue and red points corresponding to $\delta^{\xi}_+$ and $\delta^{\xi}_-$. Here, only the positive root is physical since $\delta^{\xi}_+ \rightarrow 0$ when $\mathbf{y}_- \rightarrow 0$ according to \textbf{Lemma} \ref{lemma 3}. Fig. \ref{fig 3}(b) shows the imaginary and real part of the first-order ($\delta^{\xi}_1$) vs higher-order ($\delta^{\xi}_+$) flight time estimate as $y_1$ is varied by keeping the norm fixed at $||\mathbf{y}_-|| = 0.00450243$. Impacts only occur when the imaginary part of $\delta$ vanishes and are given by the points lying on the green surface $\text{Im}(\delta) = 0$. The higher-order theory predicts that impacts will only occur for a range of values of $y_1$ as shown in Fig. \ref{fig 3}(b). On the contrary, the first-order theory predicts that impacts occur for all perturbations and does not depend on the state $\mathbf{y}_-$ during impact at $\mathbf{x}_i$. Similarly, Fig. \ref{fig 3}(c) and (d) presents a surface plot of $G(\delta, \mathbf{x}_i, \mathbf{y}_-)$ vs $\delta$ and $y_1$ and, the imaginary and real part of $\delta^{\xi}_1$ and $\delta^{\xi}_+$ vs $y_1$ as the second component of $\mathbf{y}_-$ is kept fixed at $-0.00115247$. Once again, solutions for impacts are given by the curve formed due to the intersection of the surface $G(\delta, \mathbf{x}_i, \mathbf{y}_-)$ with the blue plane $G(\delta^{\xi}_+, \mathbf{x}_i, \mathbf{y}_-) = 0$; see Fig. \ref{fig 3}(c). The locus of the points is given by the roots of $G(\delta, \mathbf{x}_i, \mathbf{y}_-) = 0$ (shown as blue and red points) and only the positive root is physical. Fig. \ref{fig 3}(d) shows that impacts only occur when $y_1$ is smaller than some critical value below which $\text{Im}(\delta^{\xi}_+) = 0$ is ensured. This particular case of Fig. \ref{fig 3} for $\sigma = -0.11$, where no impacts can occur, is validated by a direct numerical simulation presented in Figs. \ref{fig 5}(c) and (d). 

\begin{figure}[!]
    \centering
    \includegraphics[scale = 0.8]{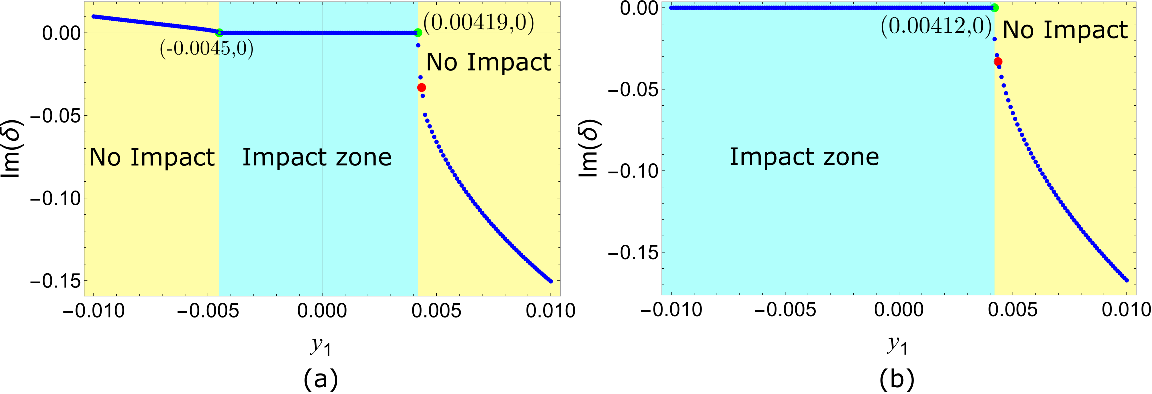}
    \caption{Imaginary part of $\delta^{\xi}_+$ vs $y_1$ with (a) norm of $\mathbf{y}_-$ fixed at $||\mathbf{y}_-|| = 0.00450243$ and (b) second component of $\mathbf{y}_-$ fixed at $y_2 = -0.00115247$. The cyan region corresponds to the impact zone where perturbations reach the discontinuity boundary provided they satisfy (a) $-0.0045 \leq y_1 \leq 0.00419$ and (b) $||\mathbf{y}_-|| \leq 0.00412$. Results correspond to impact states $\mathbf{x}_i = [-0.11, -0.0577068]$ and $\mathbf{y}_- = [0.00435243, -0.00115247]$ at $t_i = 3489.83$ with system parameters $\xi = 2.0$, $\omega = 1.8$ and $\sigma = -0.11$.}
    \label{fig 4}
\end{figure}

\begin{figure}[h!]
    \centering
    \includegraphics[scale = 0.8]{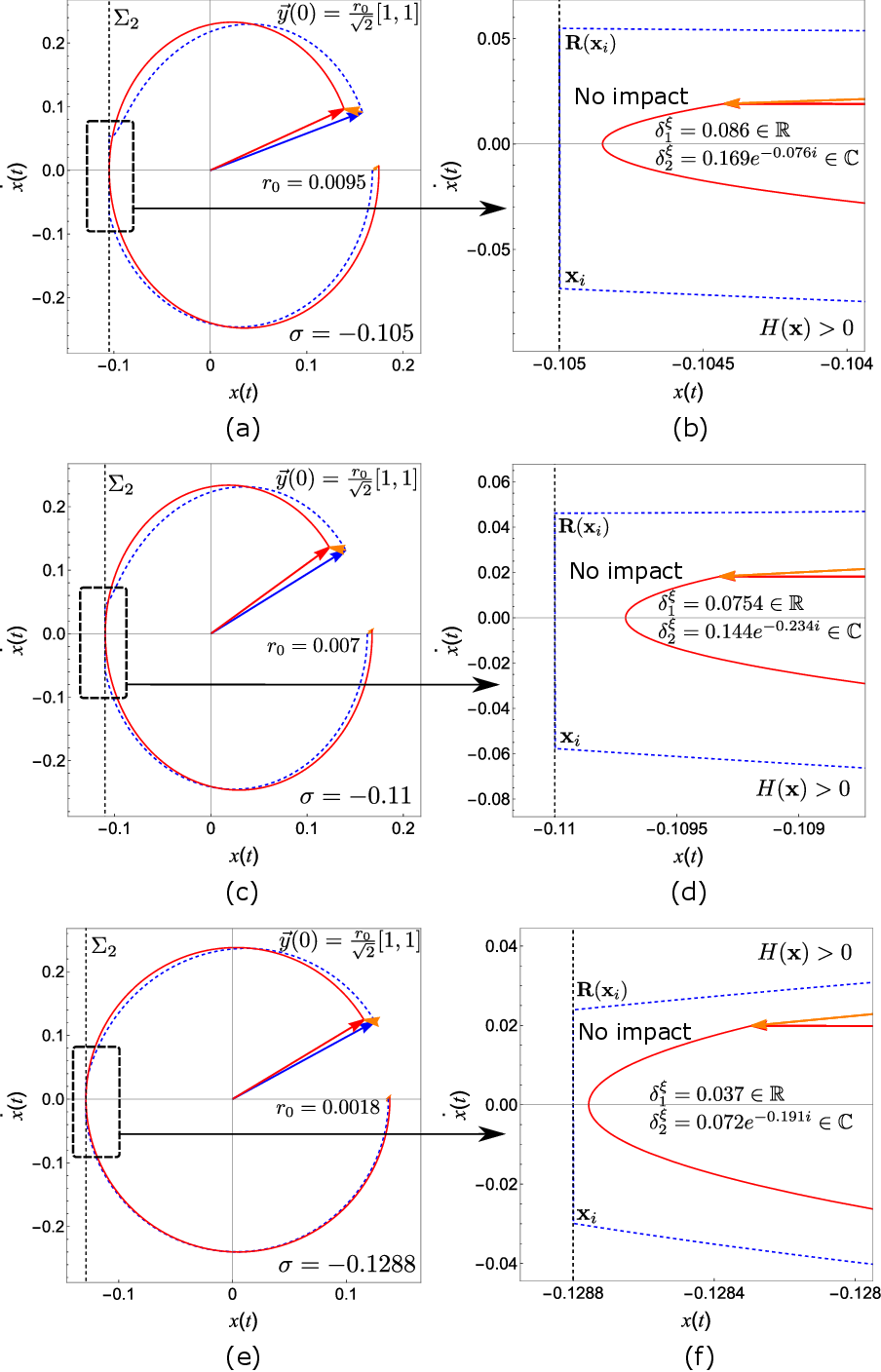}
    \caption{Phase-portraits demonstrating perturbed orbits not impacting the discontinuity boundary as predicted by the higher-order TDM. Nominal and perturbed trajectories for three different perturbations and barrier distance: $r_0 = 0.0095$ and $\sigma = -0.105$ in (a), $r_0 = 0.007$ and $\sigma = -0.11$ in (c), and, $r_0 = 0.0018$ and $\sigma = -0.1288$ in (e) with their corresponding zoomed-phase portraits in (b), (d) and (f). Trajectories $\mathbf{x}$ and $\mathbf{x} + \mathbf{y}$ are shown in blue and red with initial perturbation $\mathbf{y} = r_0/\sqrt{2}[1,1]$ and parameters set to $\xi = 2.0$, $\omega = 1.8$ and $r = 0.8$.}
    \label{fig 5}
\end{figure}

%Surface plot of  and, (b) Scatter plot showing imaginary and real parts of flight time $\delta$ vs $y_1$ with norm of $\mathbf{y}_-$ fixed at $||\mathbf{y}_-|| = 0.00450243$. (c) Surface plot of $G(\delta, \mathbf{x}_i, \mathbf{y}_1)$ vs $\delta$ and $y_1$ and, (d) Scatter plot showing imaginary and real parts of flight time $\delta$ vs $y_1$ with second component of $\mathbf{y}_-$ fixed at $y_2 = -0.00115247$. Results correspond to $\xi = 2.0$, $\omega = 1.8$, barrier placed at $\sigma = -0.11$ and $\mathbf{x}_i = [-0.11, -0.0577068]$ at $t=3488.19$.
The critical range within which perturbation vectors ($y_1$ and $||\mathbf{y}_-||$) should be bounded to ensure that they reach the discontinuity boundary can be determined analytically by equating the discriminant of Eq. \eqref{eq 10} to zero. This results in the inequality,
\begin{equation}\label{eq 22}
    (2(v + y_2))^2 - 8(\cos{\omega t_i} - \sigma - 2\xi v)y_1 \geq 0,
\end{equation}
where the impact state is $\mathbf{x}_i = [\sigma, v]$. For the case shown in Fig. \ref{fig 3} with barrier located at $\sigma = -0.11$, the critical ranges of $\mathbf{y}_-$ are shown in Fig. \ref{fig 4}. Fig. \ref{fig 4}(a) represents the imaginary part of $\delta^{\xi}_+$ vs $y_1$ with norm fixed at $||\mathbf{y}_-|| = 0.00450243$. From Eq. \eqref{eq 22}, it can be found that impacts only occur when $y_1$ ranges between $-0.0045 \leq y_1 \leq 0.00419$. The cyan region (impact zone) in Fig. \ref{fig 4}(a) corresponds to the perturbation vectors that will reach the discontinuity boundary, while orbits in the yellow region do not reach the barrier and continue evolving in the phase space where $H(\mathbf{x})>0$. Similarly, Fig. \ref{fig 4}(b) plots the imaginary part of $\delta^{\xi}_+$ vs $y_1$ with $y_2$ fixed at $y_2 = -0.00115247$. Note here that as $y_1$ is varied while keeping $y_2$ fixed, the norm also changes. The inequality of Eq. \eqref{eq 22} predicts that perturbation vectors with norm $||\mathbf{y}_-|| \leq 0.00412$ can only reach the discontinuity boundary, shown by the cyan region (impact zone) in Fig. \ref{fig 4}(b). These higher-order analytical predictions are validated next.

Fig. \ref{fig 5}(a), Fig. \ref{fig 5}(c) and Fig. \ref{fig 5}(e) are the phase-portraits of two nearby trajectories $\mathbf{x}$ and $\mathbf{x} + \mathbf{y}$ where $\mathbf{y}(0) = r_0/\sqrt{2}[1,1]$ undergoing impact at $\sigma = -0.105$, $\sigma = -0.11$ and $-0.1288$ respectively. The region near the discontinuity boundary is magnified in Figs. \ref{fig 5}(b), (d), and (f). Transient effects are disregarded by rejecting the initial $1000$ impacts to ensure the formation of a periodic orbit. The initial separation, $||\mathbf{y}(t_0)||$, between the trajectories is taken as $r_0 = 0.0095$ in Fig. \ref{fig 5}(a), $r_0 = 0.007$ in Fig. \ref{fig 5}(c) and $0.0018$ in Fig. \ref{fig 5}(e) with $\xi = 2.0$, $\omega = 1.8$ and $r = 0.8$. In all three cases, while the trajectory $\mathbf{x}$ undergoes an impact at $\sigma$, the perturbed trajectory $\mathbf{x} + \mathbf{y}$ misses the barrier as verified by the respective magnified phase-portraits of the direct numerical solution shown in Figs. \ref{fig 5}(b), (d) and (f). However, a linearized approximation using $\mathcal{O}(1)$ incorrectly predicts an impact after $\delta^{\xi}_1 = 0.086$, $\delta^{\xi}_1 = 0.0754$ and $\delta^{\xi}_1=0.037$. Hence, the first-order saltation matrix yields an incorrect mapping of the perturbed state from $\mathbf{x}_2$ to $\mathbf{x}_3$. The $\mathcal{O}(2)$ terms in Eqs. \eqref{eq 21} returns an imaginary root for all three cases, leading to a logical conclusion that there is no interaction with $\Sigma_2$. This is because no real root exists which satisfies $H(\mathbf{x}_2) = 0$ for the chosen impact state $\mathbf{x}_i$. When compared with the direct numerical simulations shown in Figs. \ref{fig 5}(b), (d), (f), the perturbed trajectory, governed by the variational equation, is observed to miss the discontinuity barrier as predicted by the higher-order flight time $\delta^{\xi}_+ \in \mathbb{C}$. Thus, the higher-order approximation of flight time correctly predicts the behaviour of perturbed orbits near a barrier, while the first-order TDM fails to do so. 

\begin{figure}[!]
    \centering
    \includegraphics[scale = 0.8]{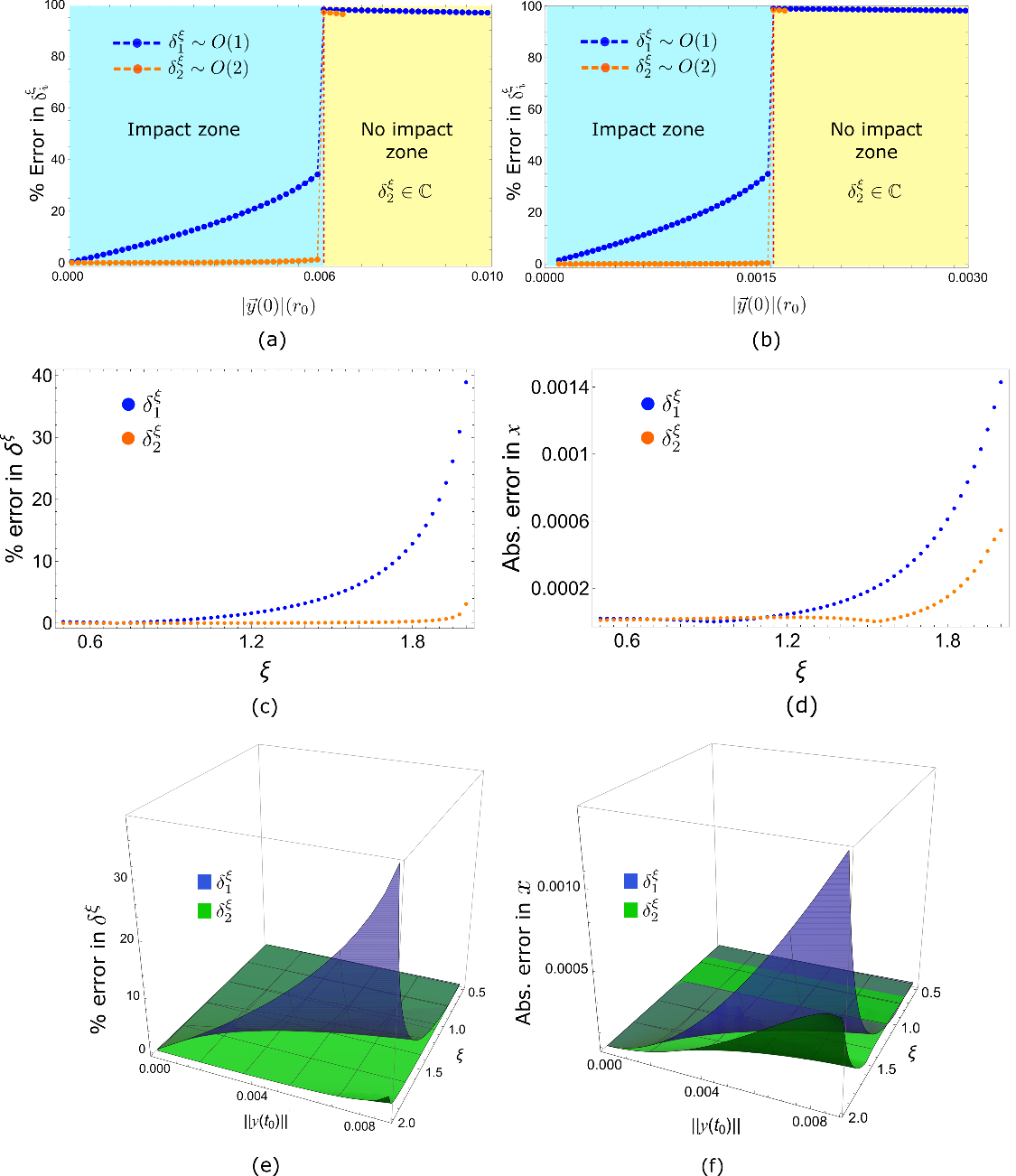}
    \caption{Comparison of first-order $\delta^{\xi}_1$ and higher-order flight time $\delta^{\xi}_+$ while estimating $\mathbf{x}_3$. Figs. (a) - (b) show percentage error in $\delta^{\xi}_i$ vs initial perturbation magnitude $r_0$ for barrier placed at (a) $\sigma = -0.11$ (b) $\sigma = -0.1288$. Cyan and yellow regions are no impact and impact zones when compared to exact numerical results. (c) The percentage error in $\delta^{\xi}$, (d) absolute error in position or first-component of state $\mathbf{x}_3$ for varying damping parameter $\xi$ and initial separation $||\mathbf{y}(t_0)|| = 0.0089$, (e) percentage error in $\delta^{\xi}$ and, (f) absolute error in position $x$ of state $\mathbf{x}_3$ for varying damping parameter $\xi$ and initial separation $r_0 = ||\mathbf{y}(t_0)||$. Here, in Figs. (c) - (f), $\omega = 1.8$, $r = 0.8$ and barrier at $\sigma = -0.105$.}
    \label{fig 6}
\end{figure}

Next, the percentage error is compared between the first-order and higher-order approaches. In Fig. \ref{fig 6}, the percentage error in $\delta^{\xi}$ is shown as the initial separation between the trajectories is varied when the barrier is placed at (a) $\sigma = -0.11$ and (b)$\sigma = -0.1288$. The cyan and yellow regions correspond to the impact and no impact zones, respectively. In the impact zone, $\delta^{\xi}_+$ gives a better approximation of $\delta$ in comparison to $\delta^{\xi}_1$ as the separation $r_0$ increases. Improvements up to $40\%$ were obtained while using the higher-order terms in $\delta^{\xi}_+$. In the no-impact zone, $\delta^{\xi}_+$ returned imaginary values, while the first-order $\delta^{\xi}_1$ incorrectly predicted impacts with the barrier. Note that the perturbation vectors have a norm in the range $\mathcal{O}(10^{-3})$. This is a major improvement over the linearized TDM, illustrating that the higher-order can correctly predict the evolution of perturbed orbits near the discontinuity boundary. Figs. \ref{fig 6}(c)-(f) highlights the dependence of crucial parameters like damping $\xi$ in the higher-order flight time $\delta^{\xi}_+$ which is absent in the first-order flight time $\delta^{\xi}_1$; compare Eqs. \eqref{eq 21}. Here, the barrier is located at $\sigma = -0.105$. Fig. \ref{fig 6}(c) compares the percentage error in flight time $\delta^{\xi}$ estimated by the first-order and higher-order expressions of Eqs. \eqref{eq 21} for varying damping with $\omega = 1.8$, $r = 0.8$. Fig. \ref{fig 6}(d) compares the absolute error in the estimation of the position of the oscillator (the first component of the mapped state $\mathbf{x}_3$) after a time elapse $t_i + \delta^{\xi}$ for varying damping. For both the Figs. \ref{fig 6}(c) and (d), the flight time $\delta^{\xi}_1$ and $\delta^{\xi}_+$ has been estimated between two trajectories with an initial separation of $r_0 = ||\mathbf{y}(t_0)|| = 0.0089$. Similarly, Figs. \ref{fig 6}(e) and (f) compare the percentage error of $\delta^{\xi}$ and the absolute error in the position of the mapped state $\mathbf{x}_3$ by varying the damping parameters and for various initial separation $r_0 = ||\mathbf{y}(t_0)||$. Results show that $\delta^{\xi}_+$ (see Eqs. \eqref{eq 21}) considers the damping parameter $\xi$, which improves the accuracy of the estimated flight times and mapped states. On the contrary, the first-order $\delta^{\xi}_1$ does not depend on crucial system parameters like damping, which reduces the accuracy of the conventional first-order approaches. Further, additional terms like $\xi$ and $\cos{\omega t_i}$ in the denominator which are not present in $\delta^{\xi}_1$ results in an overestimation of the mapped state $\mathbf{x}_3$ as demonstrated next.

\begin{figure}[!]
    \centering
    \includegraphics[width = 0.70\linewidth]{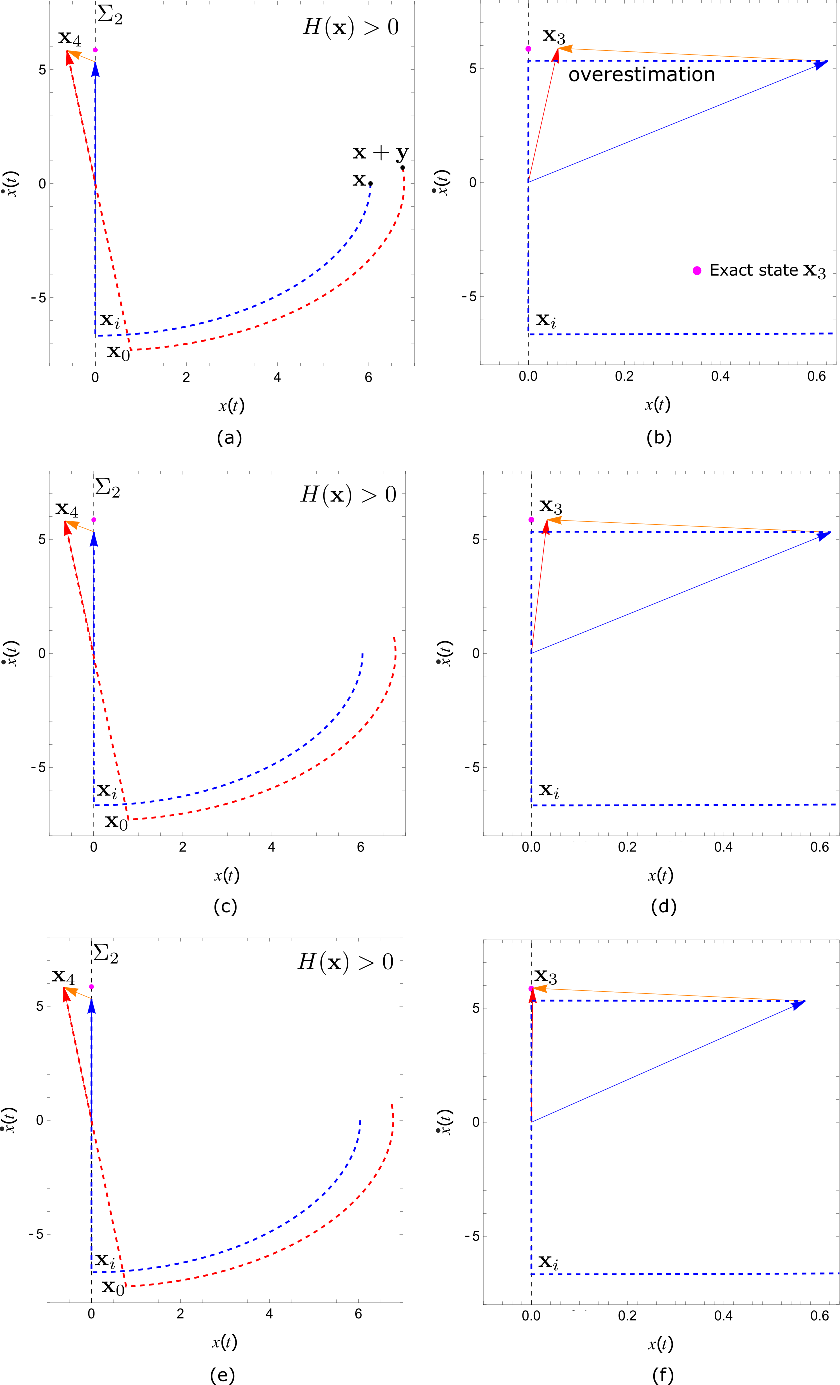}
    \caption{Phase portrait of two trajectories $\mathbf{x}$, shown in blue, and $\mathbf{x} + \mathbf{y}$, shown as red dashed lines, are separated initially by a perturbation $\lVert\mathbf{y}\rVert = 1.0$. The magenta dot depicts the numerically obtained mapped state $\mathbf{x}_3$ that $\mathbf{x}+\mathbf{y}$ gets mapped to after $t = t_i + \delta$. The discontinuity boundary is shown as a black dashed line. Phase portraits showing the estimated mapped state $\mathbf{x}_3$ at time $t = t_i + \delta$ using: $\mathcal{O}(1)$ in $\delta$ and $\mathcal{O}(1)$ in saltation terms (a) before impact, (b) after impact; $\mathcal{O}(1)$ in $\delta$ and $\mathcal{O}(2)$ in saltation terms (c) before impact, (d) after impact; $\mathcal{O}(2)$ in $\delta$ and $\mathcal{O}(1)$ in saltation terms (e) before impact, (f) after impact. Results correspond to $\omega = 2.0$, $\xi = 0.0$, $r = 0.8$ and $\sigma = 0.0$.}
    \label{fig 7}
\end{figure}

The first-order saltation matrix is compared when using the first-order versus the higher-order flight time for the case of an undamped impact oscillator. The expressions in Eqs. \eqref{eq 23} corresponds to various approximations of the flight time $\delta$ incorporating higher-order correction terms. The entity $\delta_1$ in Eq. \eqref{eq 23}(a) depicts the first-order approximation of Eq. \eqref{eq 10}, \textit{i.e.,} Eq. \eqref{eq 16}. The variables $\delta_2$ and $\delta_3$ in Eq. \eqref{eq 23}(b) and (c) are obtained from Eq. \eqref{eq 21} by expanding the discriminant up to $\mathcal{O}(1)$ and $\mathcal{O}(2)$ respectively. Meanwhile, $\delta_4$ is the positive root of Eq. \eqref{eq 21}. 
\begin{subequations} \label{eq 23}
    \begin{align}
        \delta_1 &= -\dfrac{y_1}{v_{-}}, \\
        \delta_2 &= -\dfrac{y_1}{v_{-} + y_2}, \\
        \delta_3 &= -\dfrac{y_1}{v_{-} + y_2} - \dfrac{(-\sigma + \cos{(\omega t_i)})y_1^2}{2(v_{-} + y_2)^3}, \\
        \delta_4 &= -\dfrac{v_- + y_2}{-\sigma + \cos{(\omega t_i)}} \\ 
        & \qquad +\dfrac{v_- + y_2}{-\sigma + \cos{(\omega t_i)}} \sqrt{1-2\dfrac{(-\sigma + \cos{(\omega t_i)})y_1}{(v_- + y_2)^2}}. \nonumber
    \end{align}
\end{subequations} 
\noindent To demonstrate the accuracy of the higher-order TDM, the phase-portrait of two nearby trajectories is shown in Fig. \ref{fig 7}. The orbits are initially separated by a perturbation vector with a unit norm $\lVert \mathbf{y} \rVert = 1$. A unit norm is considered as the monodromy matrix is constructed from a unit sphere of perturbed vectors, ensuring that the initial fundamental solution matrix is an identity matrix. The trajectory shown in blue corresponds to a steady state orbit for the impact oscillator with $\omega = 2.0$, $\sigma = 0.0$ and $r = 0.8$ while the perturbed orbit is shown in red. In principle, from the instant of impact to the instant when $t = t_i + \delta$, the perturbed trajectory should reach the state $\mathbf{x}_3$ on the discontinuity boundary if correctly mapped. Fig. \ref{fig 7}(a) shows the trajectories at the instant of impact after the application of the TDM, when mapped using $\mathcal{O}(1)$ time difference and TDM ({\it i.e.}, Eq. \eqref{eq 16} and Eq. \eqref{eq 19}), are shown. Fig. \ref{fig 7}(b) corresponds to the trajectories after the application of TDM after elapsed time $\delta$ \textit{i.e.}, when $t = t_i + \delta$. The mapped perturbed vector does not lie on the discontinuity boundary due to overestimation of $\mathbf{x}_3$, depicting the inaccuracy of the first-order mapping. An improvement in mapping is observed in Fig. \ref{fig 7}(c) - (d) where the mapping is done using an $\mathcal{O}(1)$ time difference and an $\mathcal{O}(2)$ TDM, {\it i.e.} using Eq. \eqref{eq 16} and Eq. \eqref{eq 11}. This further reduces when one uses a $\mathcal{O}(2)$ approximation in $\delta$ and a $\mathcal{O}(1)$ TDM using Eq. \eqref{eq 9} and Eq. \eqref{eq 19}; See Figs. \ref{fig 7}(e) - (f). It can thus be inferred that the higher-order $\delta_+$ provides a significant improvement since the correction terms avoid an overestimation of the state $\mathbf{x}_3$. This ensures that the flight time does not diverge, especially for low-velocity impacts.
%A mapping $\mathcal{O}(2)$ in $\delta$ as well as the saltation may lead to a more accurate estimate, but this yields an expression higher than $\mathcal{O}(2)$ and is not presented here.

\begin{figure}[!]
    \centering
    \includegraphics[scale = 0.8]{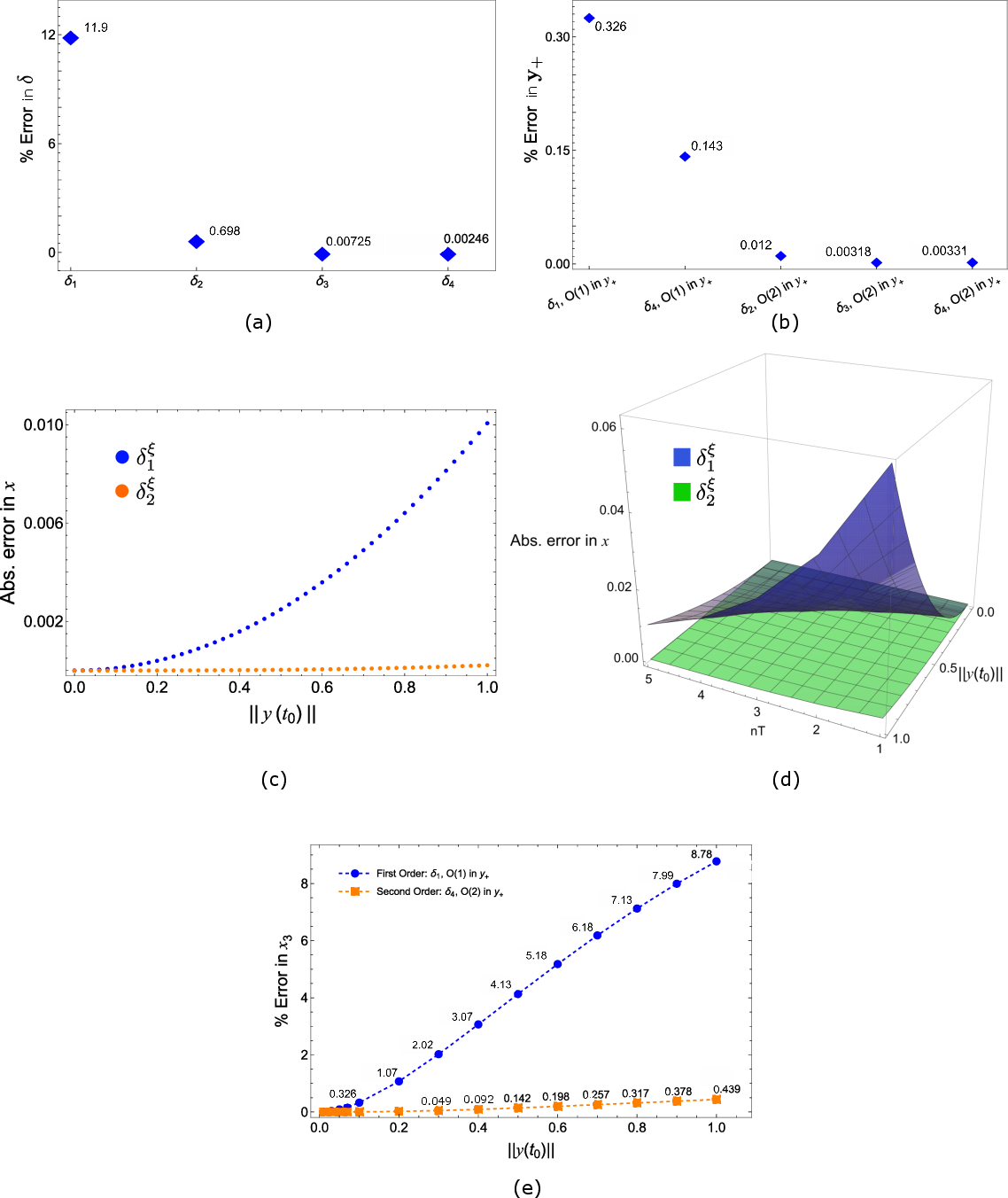}
    \caption{Comparison of percentage error between first-order and higher-order TDM for system parameters $\omega = 2.0$, $r = 0.8$ and $\sigma = 0.0$. Shown are: percentage errors in (a) flight time approximations of $\delta$ from Eq. \eqref{eq 23} and (b) mapped perturbation $\mathbf{y}_+$, absolute errors in position component of $\mathbf{x}_3$ after (c) 5 impact cycles and (d) for varying impacting cycles, and percentage errors in (e) velocity component of mapped perturbation $\mathbf{x}_3$ as initial separation $\|\mathbf{y}(0)\|$ varies.}
    \label{fig 8}
\end{figure}

Next, a comparison of the mapped state $\mathbf{x}_3$ between first and higher-order TDM is presented after multiple impacts over a long time of integration. Fig. \ref{fig 8}(a) shows the percentage error in various approximations of $\delta$ using Eqs. \eqref{eq 23} while Fig. \ref{fig 8}(b) shows the percentage error in the numerically estimated mapped state $\mathbf{y}_+$. Figs. \ref{fig 8}(c) and (d) compare the errors in position (first component of $\mathbf{x}_3$) estimated after a long-time integration of the impact oscillator with $\omega = 2.0$, $r = 0.8$ and $\sigma = 0.0$, as the separation between trajectories is varied. Fig. \ref{fig 8}(c) compares the absolute error in position after implementing the first-order and higher-order TDM (Eqs. \ref{eq 19} and Eqs. \ref{eq 11}) after $5$ impacts. Here, the impact oscillator is integrated up to $n = 5$ time periods ($T = 2\pi/\omega$). Similarly, Fig. \ref{fig 8}(d) compares the absolute error in the position component of $\mathbf{x}_3$ while varying the initial separation $||\mathbf{y}(t_0)||$ and applying the first-order and higher-order TDM for $n = 1$ to $5$ impacts with the discontinuity boundary. Results show that the higher-order flight time $\delta^{\xi}_+$ accurately predicts the state $\mathbf{x}_3$ after all cumulative impacts in comparison to the first-order $\delta^{\xi}_1$. Fig. \ref{fig 8}(e) compares the percentage error in $\dot{x}$ or velocity component of $\mathbf{x}_3$ between the first and higher-order TDM. The error in mapping is greatly minimised at every impact occurrence and is of the order of $10^{-3}$ when a combination of $\delta_4$ (Eq. \eqref{eq 23}(d)) and $\mathbf{x}_4$ (Eq. \eqref{eq 11}) is implemented.
%\begin{figure}[!]
%    \centering
%    \includegraphics[scale = 0.65]{Figures/percentErrorTime.eps}
%    \caption{Percent errors in prediction of flight time \textit{i.e.}, $\delta_i$s as defined in Eq. \eqref{eq 19}, in comparison to the numerically obtained flight times. The computations are carried out for forcing frequency, $\omega = 2.0$ with a barrier at $\sigma = 0.0$ and $r = 0.8$. The initial separation between trajectories is $\lVert\mathbf{y}\rVert = 0.1$.}
%    \label{}
%\end{figure}

%\begin{figure}[!]
%    \centering
%    \includegraphics[scale = 0.65]{Figures/percentErrorVelocity.eps}
%    \caption{Comparison of errors in predicted $\mathbf{y}_{+}$ for different combinations of $\delta_i$s and $\mathcal{O}(1)$, $\mathcal{O}(2)$ mapping of $\mathbf{y}_{+}$ with respect to the numerical obtained exact values. Results are shown for $\omega = 2.0$, $r = 0.8$ and $\sigma = 0.0$. Initial separation between trajectories is $\lVert\mathbf{y}\rVert = 0.1$.}
%    \label{}
%\end{figure}
An improvement of the discontinuity mapping has greater consequences. Any hybrid dynamical system attains a steady state after multiple impacts. If a first-order framework is used, these errors will add up, leading to a completely different trajectory. This can result in an incorrect prediction of the stability of a non-smooth limit cycle. Additionally, the errors introduced when using a first-order saltation matrix increase as the separation between orbits increases, as shown in Fig. \ref{fig 8}. Further, the higher-order TDM can also predict impact occurrence, provided the discriminant in Eq. \eqref{eq 10} is positive. The first-order TDM cannot capture this, resulting in a mapping of orbits ($\mathbf{x} + \mathbf{y}$) local to the impact state $\mathbf{x}$, although it cannot reach the discontinuity boundary. To further confirm that the higher-order TDM works during several impacts within one periodic orbit, an impact oscillator with two discontinuity boundaries is considered next.

\subsection{Pair impact oscillator}

\begin{figure}[!]
    \centering
    \includegraphics[scale = 1.25]{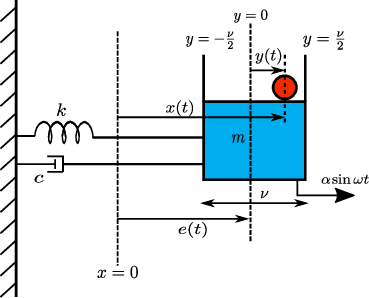}
    \caption{Point mass placed on a cart excited by harmonic forcing. Impact occurs when $y = \pm \nu/2$.}
    \label{fig 9}
\end{figure}

\noindent A mechanical oscillator undergoing impacts with two rigid undeformable barriers is considered \cite{han1995chaotic}. The schematic of this pair-impact oscillator is shown in Fig. \ref{fig 9}. The model comprises a typical mass-cart system executing periodic motion, where $x(t)$ and $y(t)$ describe the displacement of the object with respect to the stationary frame and the frame of reference, respectively. The term $e(t) = \alpha \sin{(\omega t)}$ describes the periodic motion of the cart excited externally with frequency $\omega$. The dynamic response of the mass-cart is represented by $x(t) = y(t) + e(t)$ and can be expressed as,
\begin{equation} \label{eq 24}
    \begin{cases}
    \ddot{y} = \alpha \omega^2 \sin{(\omega t)}, \quad \ \text{if} \ |y| < \dfrac{\nu}{2},\\
    \dot{y}(t_+) = -r\dot{y}(t_-), \quad \text{if} \ |y(t_-)| = \dfrac{\nu}{2}, 
    \end{cases}
\end{equation}
where the width of the cart is expressed by $\nu$. Naturally, the motion of this point mass object will be obstructed at either wall of the cart when $y(t_-) = \pm \nu/{2}$. Here, an instantaneous velocity reversal occurs and is defined in Eq. \eqref{eq 24} where $t_-$ and $t_+$ denote instants before and after impact. As defined for the impact oscillator, $r$ represents a coefficient of restitution.

The next section verifies the accuracy of the higher-order TDM to predict the bifurcation behaviour for the two representative hybrid systems, \textit{i.e.}, the impact and the pair-impact oscillator. A numerical method to estimate the Lyapunov spectrum by incorporating the proposed higher-order TDM is presented in the next section. The proposed algorithm correctly estimates occurrences of impacts with the discontinuity boundary transversally, which the first-order saltation matrix cannot capture. Therefore, this higher-order approach can serve to estimate LEs for any generalized PWS hybrid systems correctly.

%The stability of a hybrid dynamical system was previously assessed through analytical solution of the smooth trajectories between impacts\cite{bernardo2008piecewise}. The merit of the proposed algorithm in this paper lies in the fact that it does not require any prior knowledge of the analytical solution of the system near the discontinuity boundary.

\section{Lyapunov exponents} \label{sec 4}

%\noindent The stability analysis of the impact and pair-impact oscillator, defined in section \ref{sec 3}, is demonstrated next. 
\noindent Since Lyapunov characteristic exponents (LE) measure the exponential divergence between two closely spaced trajectories \cite{benettin1980Lyapunov1,bennetin1980Lyapunov2,strogatz2018nonlinear}, it is essential to correctly map orbits in the local neighbourhood of the discontinuity boundary. It was demonstrated in section \ref{sec 3} how the first-order saltation matrix predicts an impact even though the perturbation vector does not reach the discontinuity boundary. This is resolved by implementing the higher-order TDM (see Eqs. \eqref{eq 10} and \eqref{eq 11}) that defines the behaviour of perturbation vectors near the discontinuity boundary when the dynamical state $\mathbf{x}$ of an impact oscillator encounters an impact with the rigid barrier. Orbits in the local neighbourhood of $\mathbf{x}$ only show an impact when the discriminant of the quadratic function of Eq. \eqref{eq 9} is positive. The method to calculate LEs by incorporating a higher-order TDM is discussed in \ref{app a}.

\subsection{Impact oscillator}
\begin{figure}[!]
    \includegraphics[scale = 0.8]{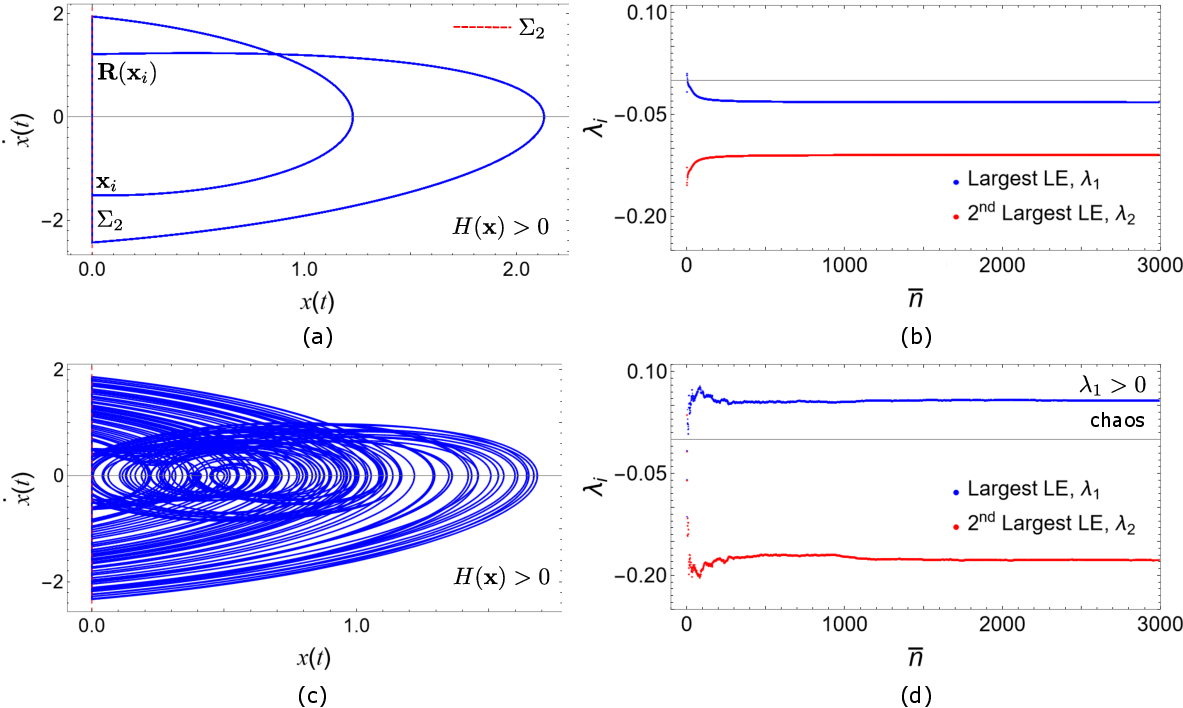}
    \caption{Phase portrait of the impact oscillator corresponding to (a) $\omega = 1.0$ (c) $\omega = 1.1$ and the respective Lyapunov spectrum for (b) $\omega = 1.0$ and (d) $\omega = 1.1$. Here $\xi = 0$, $r = 0.8$, and barrier is placed at $\sigma = 0$ (red dashed line).}
    \label{fig 10}
\end{figure}

A stability analysis is conducted on the hybrid impact oscillator described in Eq. \eqref{eq 20}. The external frequency, $\omega$, is taken as the bifurcation parameter. The exponential divergence between perturbed trajectories is calculated every $2\pi/\omega$. First, a hypersphere of radius $r_0 = 0.001$ is initialised. After one cycle of evolution, a QRD and re-scaling in $r_0$ is carried out. Figs. \ref{fig 10}(a) and (b) show the phase portraits and the LE spectrum, respectively, for $\omega = 1.0$. Similar plots for $\omega = 1.1$ with $r = 0.8$ are presented in Fig. \ref{fig 10}(c) and (d). For $\omega = 1.0$, the oscillator has a period-2 or P$2$ limit cycle, while for $\omega = 1.1$, the dynamics in the state space are chaotic. Here, the periodicity of an impact oscillator is defined as the number of times the trajectory crosses the Poincar\'{e} section $\dot{x} = 0$, before returning to the same, provided $\ddot{x} > 0$. The LE spectrum is plotted against the strobe count $\bar{n}$. The largest LE (LLE) is observed to be positive for $\omega = 1.1$, implying that perturbations diverge away, resulting in a chaotic orbit. At the instant of impact when $H(\mathbf{x}_i) = 0$, the perturbation vector $\mathbf{y}_-$ is mapped by the higher-order TDM derived in Eqs. \eqref{eq 10} and \eqref{eq 11} since $\mathbf{y}_+ = \mathbf{x}_4 - \mathbf{R}(\mathbf{x}_i)$. For the impact oscillator of Eq. \eqref{eq 20}, these mappings become,
\begin{subequations}\label{eq 25}
\begin{align}
    \delta_+ &= -\dfrac{v + y_2}{-\sigma + \cos{(\omega t_i)}} + \dfrac{v + y_2}{-\sigma + \cos{(\omega t_i)}} \sqrt{1-2\dfrac{(-\sigma + \cos{(\omega t_i)})y_1}{(v + y_2)^2}}, \\
    \mathbf{y}_+ &=  \begin{bmatrix}
                    y_1 + \delta_+ v(1+r) + \delta_+ y_2(1-r^2) + \dfrac{\delta_+^2}{2}(-\sigma + \cos{(\omega t)})(1 - r) -\delta_+^2 r^2 (-\sigma + \cos{(\omega t)}) \\
                    -r y_2 - \delta_+(-\sigma + \cos{(\omega t)})(1 + r) + \delta_+ y_1 (1 + r) + \delta_+^2(1 + r)(v + \dfrac{1}{2}\omega \sin{(\omega t)})
                  \end{bmatrix}
\end{align}
\end{subequations}
\noindent where $y_i$ are the components of the perturbed vector and $v$ is the velocity, $\dot{x}$ at the instant of impact, $t_i$.

\begin{figure}[!]
    \centering
    \includegraphics[scale = 0.86]{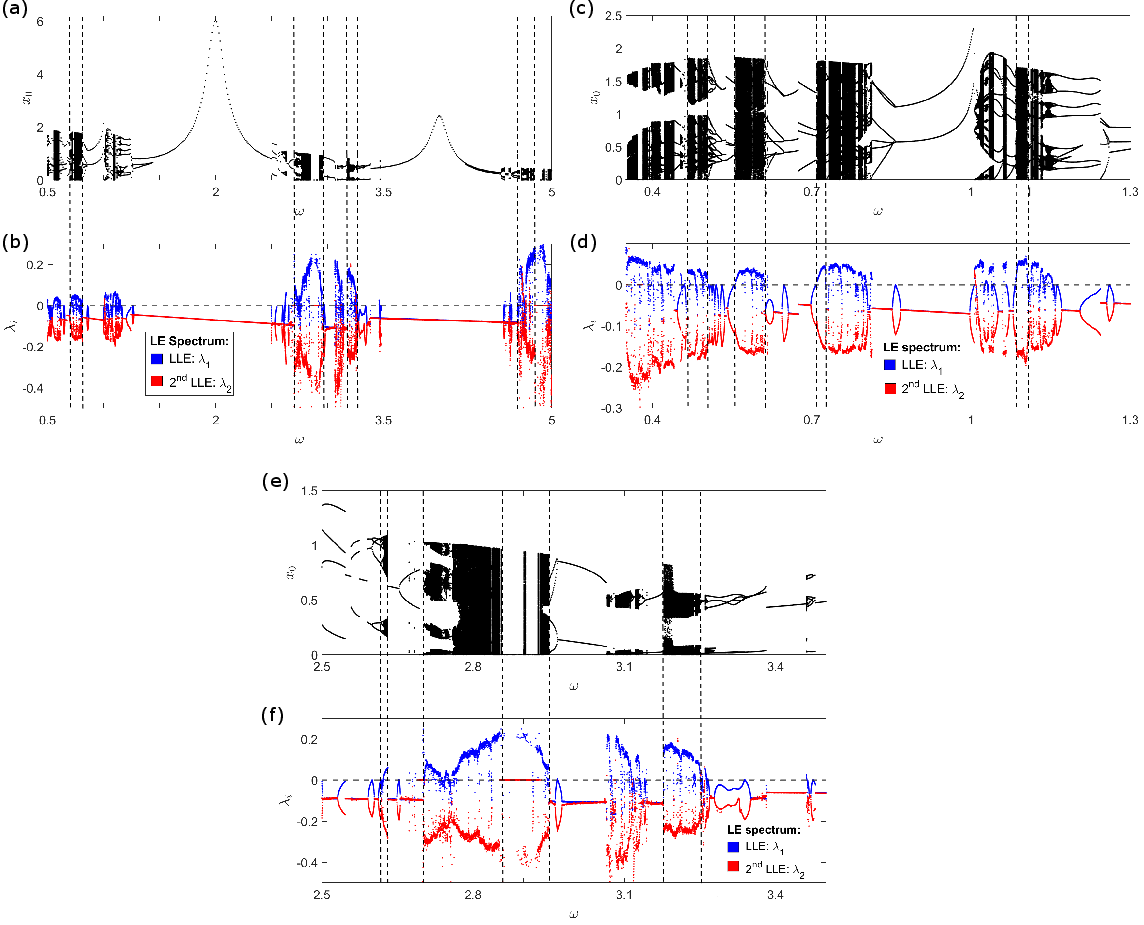}
        \caption{Bifurcation diagram showing amplitude $x_0$ vs. external frequency $\omega$ ranging between (a) $0.5 \leq \omega \leq 5.0$, (c) $0.35 \leq \omega \leq 1.3$ and (e) $2.5 \leq \omega \leq 3.5$. Corresponding LEs vs. $\omega$ ranging between (b) $0.5 \leq \omega \leq 5.0$, (d) $0.35 \leq \omega \leq 1.3$ and (f) $2.5 \leq \omega \leq 3.5$. Here, $\xi = 0$ and $r = 0.8$.}
    \label{fig 11}
\end{figure}

Fig. \ref{fig 11}(a) is a bifurcation diagram of the oscillator amplitude $x_0$ plotted against the external frequency $\omega$. The amplitude $x_0$ corresponds to the state when the orbit intersects the Poincar\'e section $\dot{x} = 0$. The integration is performed for a total of $6000$ impacts, and amplitudes for the first $3000$ impacts are discarded to eliminate any transient effects. Fig. \ref{fig 11}(b) shows the corresponding LE spectrum plotted against $\omega$ ranging between $0.5 \leq \omega \leq 5.0$. Results show that the LE spectrum is in agreement with the bifurcation diagram. Positive values of LE correspond to chaotic orbits for a chosen $\omega$ while the LLE becomes zero at the parameter value where a DIB is observed in the bifurcation diagram. Dashed vertical lines are provided in the figure for comparing the bifurcation diagram with the corresponding LEs to highlight some of the aperiodic chaotic solutions. Similarly, Figs. \ref{fig 11}(c) and (e) depict the bifurcation diagrams while Figs. \ref{fig 11} (d) and (f) are the corresponding LE spectrum for $\omega$ ranging between $0.35 \leq \omega \leq 1.3$ and $2.5 \leq \omega \leq 3.5$, respectively. It is observed that the calculated LE spectra depict the true underlying behaviour of the steady states.
%\begin{figure}[!]
%    \centering
%        \includegraphics[scale = 0.4]{Figures/ImpactLESpectrumZoom1.eps}
%        \caption{(a) Amplitude $x_0$ of the impact oscillator after 6000 impacts when $\dot{x} = 0$. (b) LE spectrum for $0.35 \leq \omega \leq 1.3$ and $r = 0.8$.}    
%    \label{fig 12}
%\end{figure}
%\begin{figure}[!]
%    \centering
%        \includegraphics[scale = 0.4]{Figures/ImpactLESpectrumZoom2.eps}
%        \caption{(a) Amplitude $x_0$ of the impact oscillator at steady state when $\dot{x} = 0$, after 6000 impacts. (b) The corresponding LE spectrum. Here, $2.5 \leq \omega \leq 3.5$ and $r = 0.8$.}
%    \label{fig 13}
%\end{figure}

\subsection{Pair impact oscillator}
\begin{figure}[!]
    \includegraphics[scale = 0.8]{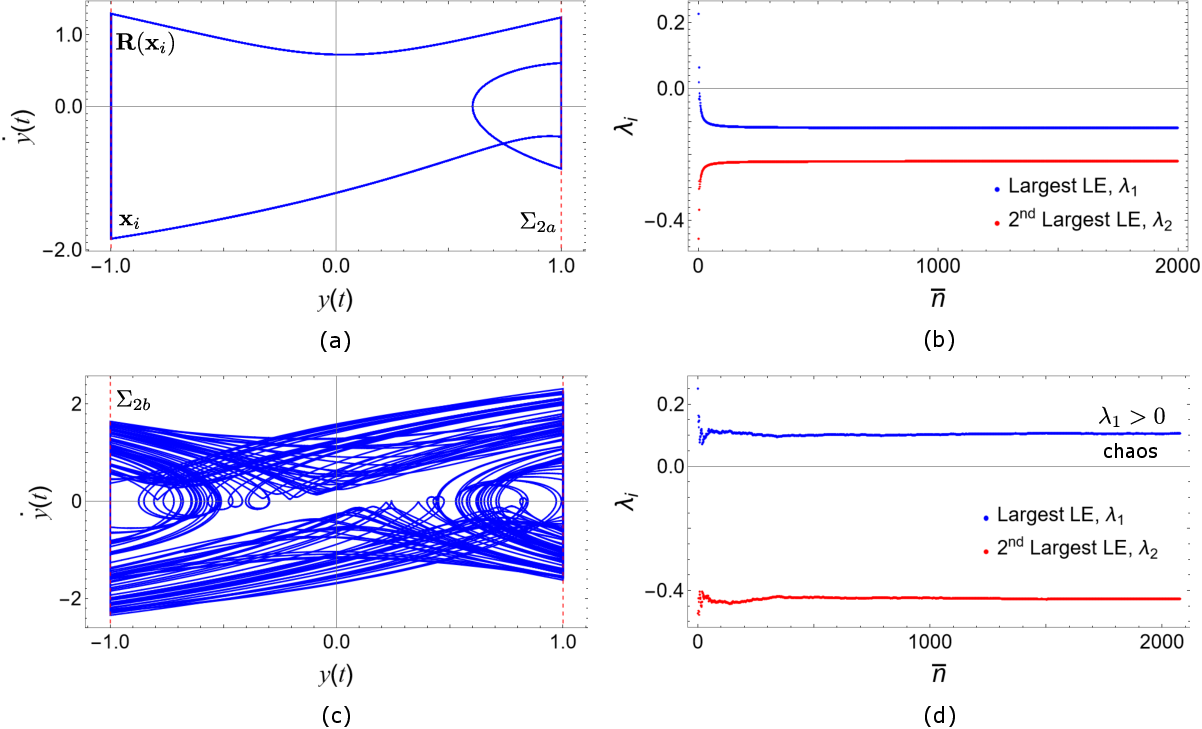}
    \caption{Phase portrait of the pair impact oscillator corresponding to (a) $\alpha = 1.0$ (c) $\alpha = 1.5$ and the respective Lyapunov spectrum for (b) $\alpha = 1.0$ and (d) $\alpha = 1.5$. Here $r = 0.7$, and barrier is placed at $\Sigma_{2a} : y = 1.0$ and $\Sigma_{2b} : y = -1.0$ (red dashed line).}
    \label{fig 12}
\end{figure}

\noindent Stability analysis of the pair-impact oscillator described in Eq. \eqref{eq 24} is carried out next. The cart has a width of $\nu = 2.0$, oscillating with a frequency of $\omega = 1.0$. The point mass moves freely on this cart unless it impacts the cart wall at $\nu = \pm \nu/2$. At this instant of impact, there is an instantaneous reversal of velocity with a coefficient of restitution $r = 0.7$ defined in Eq. \eqref{eq 24}. Figs. \ref{fig 12}(a) and (c) show the state space trajectories while Figs. \ref{fig 12}(b) and (d) are the LE spectrum for two cases of $\alpha = 1.0$ and $\alpha = 1.5$. For $\alpha = 1.0$, the orbit is stable with negative LEs, while $\alpha = 1.5$ results in a chaotic orbit with a positive LLE. For this system, the mapping is defined by Eq. \eqref{eq 26}. Here, $y_i$ are the components of the perturbed vector, and as before, $v$ is the velocity $\dot{x} = 0$ at the instant of impact, $t_i$.
\begin{subequations}\label{eq 26}
\begin{align}
    \delta_+ &= -\dfrac{v + y_2}{\alpha \omega^2 \sin{(\omega t)}} + \dfrac{v + y_2}{\alpha \omega^2 \sin{(\omega t)}} \sqrt{1-2\dfrac{(\alpha \omega^2 \sin{(\omega t)})y_1}{(v + y_2)^2}}, \\
    \mathbf{y}_+ &= \begin{bmatrix}
                    y_1 + \delta_+ v(1 + r) + \delta_+ y_2 (1 - r^2) + \dfrac{\delta_+^2}{2} \alpha \omega^2 \sin{(\omega t)} (1 - r) - \delta_+^2 r^2 \alpha \omega^2 \sin{(\omega t)} \\
                    -r y_2 - \delta_+ \alpha \omega^2 \sin{(\omega t)} (1 + r) - \dfrac{\delta_+^2}{2} \alpha \omega^3 \cos{(\omega t)} (1 + r).
                \end{bmatrix}
\end{align} 
\end{subequations}

\begin{figure}[!]
    \centering
        \includegraphics[scale = 0.4]{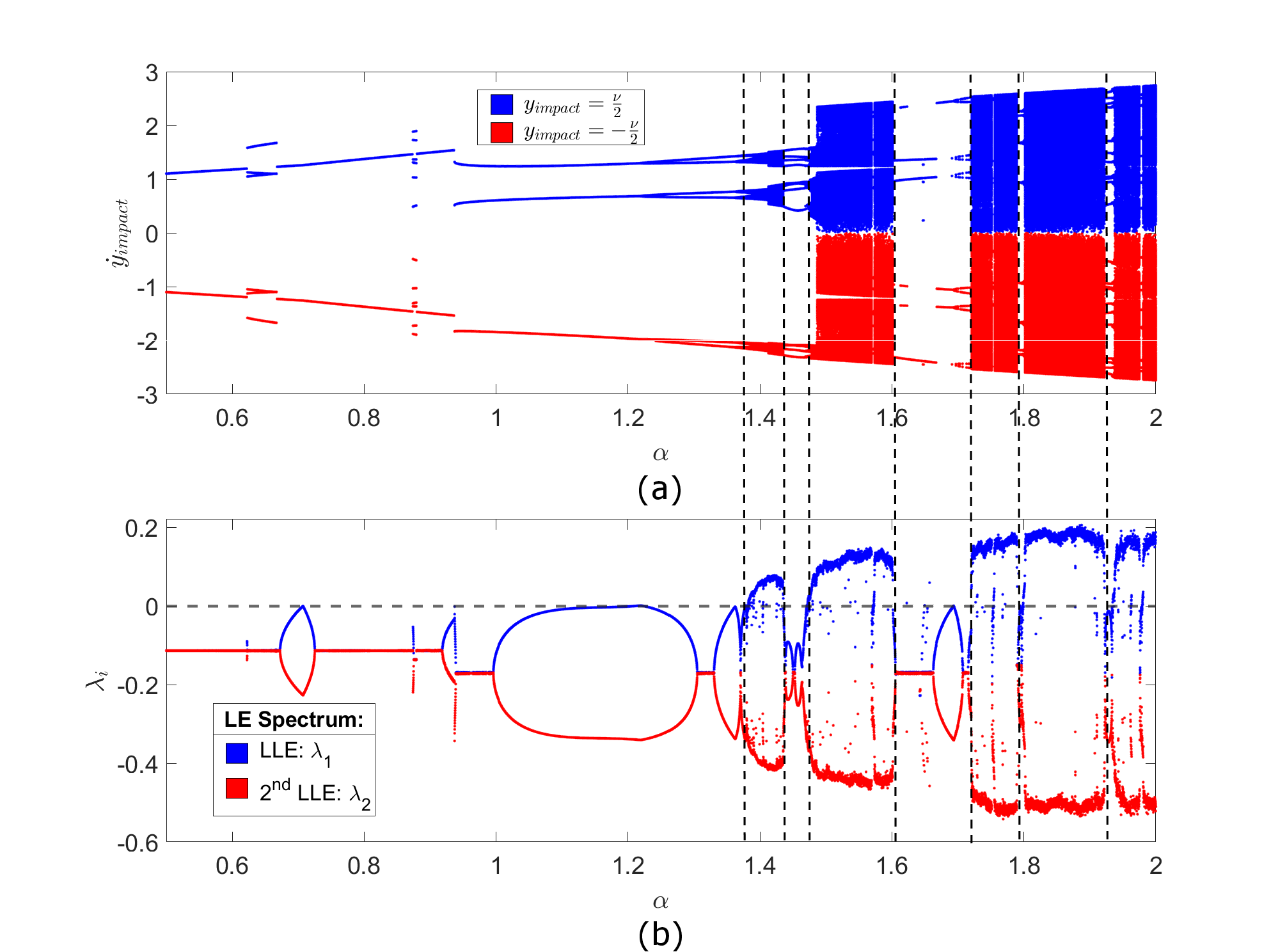}
    \caption{(a) Velocity $\dot{y}$ of steady state pair impact oscillator at instants of impact, \textit{i.e.,} $y = \pm \nu/2$, presented as a function of the forcing amplitude $\alpha$. The first 6000 impacts are discarded. (b) The corresponding LE spectrum. Here, $0.5 \leq \alpha \leq 2.0$, $\omega = 1.0$ and $r = 0.7$.}    
    \label{fig 13}
\end{figure}

Fig. \ref{fig 13}(a) shows a bifurcation diagram where the impact velocity at steady state is plotted against the corresponding oscillation amplitude $\alpha$. Here, the velocity of the point mass $\dot{y}$ is recorded at the instant of impact when $y = \pm \nu/2$ for varying $\alpha$. The bifurcation parameter ranges between $0.5 \leq \alpha \leq 2.0$. Fig. \ref{fig 13}(b) is the corresponding LE spectrum vs. $\alpha$ for the first $6000$ impacts (the first $3000$ impacts are discarded). Positive values of LLEs indicate that the underlying orbit is chaotic for the corresponding oscillation amplitude, $\alpha$. 

The next section describes a methodology to numerically calculate saltation matrices that comprise higher-order corrections terms to the TDM. Further, to validate the methodology, Floquet multipliers are calculated and compared with the respective bifurcation diagrams of the impact and pair impact oscillator.

%%%%%%%%%%%%%%%%%%%%%%%%%%%%%%%%%%%%%%%%%%%%%%%%%%%%%%%%%%%%%%%%%%%%%%%%%%%%%%%%%%%%%%%%%%%%%%%%%%%%%%%%%%%%%%%%%%%%%

\section{Higher-order saltation matrix} \label{sec 5}

\noindent The higher-order TDM derived in section \ref{sec 2} gives rise to a new challenge. Since the TDM in Eq. \eqref{eq 11} comprises terms that are proportional to $\delta_+^2$, $\delta_+ \mathbf{y}_-$ and $\mathbf{y}_-\cdot\mathbf{y}_-$, a closed-form analytical matrix transformation between $\mathbf{y}_-$ and $\mathbf{y}_+$ is not possible. This is because the perturbation state $\mathbf{y}_-$ cannot be factored out easily, like for the case of a linearized saltation matrix in Eq. \eqref{eq 19}. A matrix transformation between $\mathbf{y}_-$ and $\mathbf{y}_+$ (like the first-order saltation matrix) is necessary to conduct stability analysis from monodromy matrices. To circumvent this issue, a method to numerically obtain a saltation matrix that comprises higher-order correction terms to the TDM is described below.

For an $n^{th}$ order dynamical system $\mathbf{x} \in \mathbb{R}^n$, the variational equation whose evolution is governed by the Jacobian matrix (Eq. \eqref{eq 2}), is numerically integrated. The initial perturbed vectors are chosen along $n$ orthogonal directions, and the perturbation vectors are represented by $\mathbf{y}_i$. These vectors are expressed as a matrix as in Eq. (\ref{eq 27}).
\begin{align} \label{eq 27}
    \mathbf{Y}_{n \times n}(t) &= [\mathbf{y}_1 \ \mathbf{y}_2 \ \mathbf{y}_3 \ldots \mathbf{y}_n], \\
    \mathbf{Y}_{n \times n}(0) &= \mathbb{I}_{n \times n} \nonumber
\end{align}
The variational equations for each $\mathbf{y}_i$ coupled to the system $\mathbf{x}$ is integrated up to the instant of impact with the discontinuity boundary at $\mathbf{x} = \mathbf{x}_i$. The vectors before and after impact can be compactly expressed using matrices defined as $\mathbf{Y}_{-, impact}$ and $\mathbf{Y}_{+, impact}$ ; see Eq. (\ref{eq 28}).
\begin{subequations} \label{eq 28}
\begin{align}
    \mathbf{Y}_{+, impact} &= \begin{bmatrix}
                        y^{(1)}_{+, 1} & \ldots & y^{(1)}_{+, n} \\
                        \vdots & \ddots & \vdots \\
                        y^{(n)}_{+, 1} & \ldots & y^{(n)}_{+, n}
                    \end{bmatrix}_{n \times n}, \\
    \mathbf{Y}_{-, impact} &= \begin{bmatrix}
                        y^{(1)}_{-, 1} & \ldots & y^{(1)}_{-, n} \\
                        \vdots & \ddots & \vdots \\
                        y^{(n)}_{-, 1} & \ldots & y^{(n)}_{-, n}
                    \end{bmatrix}_{n \times n}
\end{align}
\end{subequations}
where $\mathbf{y}^{(i)}_j$ is the $i^{th}$ component of $j^{th}$ vector. A STM between perturbation vectors $\mathbf{y}_+$ and $\mathbf{y}_-$ at the instant of impact can be defined by,
\begin{equation} \label{eq 29}
    \mathbf{Y}_{+, impact} = \mathbf{S}_2\cdot\mathbf{Y}_{-, impact}
\end{equation}
where $\mathbf{S}_2$ is the saltation matrix. The subscript 2 denotes the evaluation of the saltation matrix by mapping $\mathbf{y}_-$ to $\mathbf{y}_+$ using the higher-order TDM defined in Eq. \eqref{eq 11}, which is the entity of interest. Alternatively, this higher-order saltation matrix can be evaluated by inverting Eq. \eqref{eq 29} as,
\begin{equation} \label{eq 30}
    \mathbf{S}_2 = \mathbf{Y}_{+, impact}\cdot\mathbf{Y}_{-, impact}^{-1}
\end{equation}
where the RHS of Eq. \ref{eq 30} is required to be numerically evaluated. The next section verifies that the higher-order saltation matrix derived above can correctly estimate occurrences of DIB in PWS systems.

%%%%%%%%%%%%%%%%%%%%%%%%%%%%%%%%%%%%%%%%%%%%%%%%%%%%%%%%%%%%%%%%%%%%%%%%%%%%%%%%%%%%%%%%%%%%%%%%%%%%%%%

\section{Floquet multipliers from the higher-order saltation matrix}\label{sec 6}

\begin{figure}[!]
    \centering
    \includegraphics[scale = 1.1]{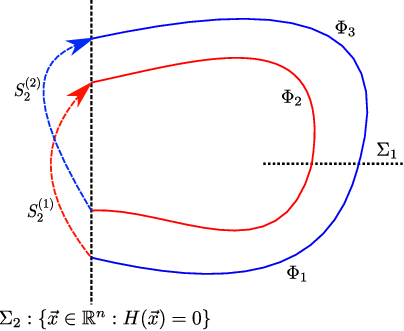}
    \caption{The schematic of a period $2$ limit cycle undergoing two instances of instantaneous reversals at the instants of impact with $\Sigma_2$, depicted as a black dashed line. The red solid line corresponds to flow governed by the STM $\Phi_2$, the red dashed line corresponds to state transition at the instant of first impact (using numerically obtained saltation matrix $S_2^{(1)}$), the blue solid line corresponds to flow governed by STM $\Phi_1$, the blue dashed line corresponds to state transition at the instant of second impact (using numerically obtained saltation matrix $S_2^{(2)}$).}
    \label{fig 14}
\end{figure}

\noindent This section presents a method to construct a monodromy matrix using the numerically obtained higher-order saltation matrix. Since eigenvalues (Floquet multipliers) of the monodromy matrix can determine the local stability of periodic solutions, they can be used as a surest test to validate the accuracy of the saltation matrix derived in Sec. \ref{sec 5}. However, the Floquet theory is applicable only to smooth and continuous dynamical systems where the Jacobian matrix is constant and periodic in time \cite{nayfeh2008applied}. The two representative impact oscillators considered in Sec. \ref{sec 3} are non-smooth and comprise instantaneous mappings during an encounter with the discontinuity boundary. Therefore, first, the periodicity of the underlying attractor is numerically determined by integrating the system until all transient effects are eliminated after a few thousand impacts with the barrier. The duration between the recurrence of state $\mathbf{x}$ across a well-defined Poincar\'{e} section post transience yields the time period of oscillations. Once the time period is calculated, the monodromy matrix is evaluated, decomposed into a product of STMs between the numerically obtained saltation matrices. The product is taken in the order of occurrence of events, making it essential to evaluate the flight times between impacts and the instances of impact. To demonstrate this effectively, a period 2 limit cycle is considered in Fig. \ref{fig 14}.

Here, $\mathbf{\Phi}_1$, $\mathbf{\Phi}_2$ and $\mathbf{\Phi}_3$ are the state transition matrices obtained from the variational equations depicting the transitions from - (1) the Poincar\'{e} section $\Sigma_1$ to the first encounter with impact surface $\Sigma_{2}$, (2) mapped $\mathbf{y}_{+}$ to second encounter with $\Sigma_{2}$, (3) mapped $\mathbf{y}_{+}$ after second encounter with $\Sigma_{2}$ to $\Sigma_{1}$ respectively, eventually completing a limit cycle with time period $T$. $\mathbf{S}^{(1)}_2$ and $\mathbf{S}^{(2)}_2$ are the saltation matrices, the closed-form expressions for which can be obtained from Eq. \eqref{eq 30}. Therefore, the monodromy matrix $\mathbf{\Phi}$ for this limit cycle with period $T$, is a resultant of the matrix multiplication given by,
\begin{equation} \label{eq 31}
    \mathbf{\Phi}(T) = \mathbf{\Phi}_3\cdot\mathbf{S}^{(2)}_2\cdot\mathbf{\Phi}_2\cdot\mathbf{S}^{(1)}_2\cdot\mathbf{\Phi}_1,
\end{equation}
where the order of matrix dot product is important, and the state transition matrices satisfy $\mathbf{\Phi}_1(0) = \mathbf{\Phi}_2(0) = \mathbf{\Phi}_3(0) = \mathbb{I}_{n \times n}$. For a hybrid dynamical system starting from an arbitrary Poincar\'{e} section $\Sigma_1$ and executing $n$ impacts with the boundary $\Sigma_{2}$ and converging to an underlying attractor with period $T$, the monodromy matrix thus takes the form represented in Eq. (\ref{eq 32}).
\begin{equation} \label{eq 32}
    \mathbf{\Phi}(T) = \mathbf{\Phi}_n\cdot \prod_{i = 1}^{n - 1} \mathbf{S}_2^{i}\cdot\mathbf{\Phi}_i 
\end{equation}
where, $\mathbf{\Phi}_{i}$ is the STM from the Poincar\'{e} section $\Sigma_1$ to the discontinuous boundary $\Sigma_{2}$. $\mathbf{S}^{i}_2$ is the saltation matrix evaluated up to $\mathcal{O}(2)$ using Eq. \eqref{eq 30} and $\mathbf{\Phi}_n$ is the final STM that takes the perturbed trajectory back to the $\Sigma_1$ within one time period $T$. To evaluate the saltation matrix $\mathbf{S}_2$ numerically, initial conditions for the set of orthogonal perturbed vectors in Eq. \eqref{eq 27} are chosen on a hypersphere of radius $r_0$. A small number is assigned to $r_0$ to ensure the perturbation vectors obey the dynamics in the local neighbourhood, governed by the Jacobian matrix. Two STMs $\mathbf{Y}_1$ and $\mathbf{Y}_2$ are defined in Eq. \eqref{eq 33} where $\mathbf{Y}_1$ evaluates the higher-order saltation matrix and $\mathbf{Y}_2$ is the STM of the flow to and from the discontinuity boundary.
\begin{align} \label{eq 33}
    \mathbf{Y}_1(0) &= r_0 \mathbb{I}_{n \times n} \text{, for evaluation of } \mathbf{S}_2^i, \\
    \mathbf{Y}_2(0) &= \mathbb{I}_{n \times n} \text{, for evaluation of } \mathbf{\Phi}_i \nonumber
\end{align}
The eigenvalues or Floquet multipliers of the monodromy matrix are numerically obtained from Eq. \eqref{eq 32}. For a dynamical system of $\mathbb{R}^n$, complex conjugate pairs of eigenvalues exist for oscillatory solutions. Magnitudes of the eigenvalues are indicators of the stability of an orbit under consideration. Stable orbits always yield eigenvalues that lie within the unit circle in the Argand plane, while those with a magnitude greater than unity denote that perturbations will diverge and result in an unstable orbit. \ref{app b} outlines the algorithm to obtain Floquet multipliers from the numerically obtained higher-order saltation matrix defined in Eq. \eqref{eq 30}.

\subsection{Impact oscillator}

\noindent In this section, Floquet multipliers of the impact oscillator (Eq. \eqref{eq 20}) are evaluated. First, the time period of the underlying attractor is obtained by integrating the system for $2000$ impacts and observing the recurring times of the return state $x$ lying on a Poincar\'e section $\dot{x} = 0$. For each chosen bifurcation parameter $\omega$, the corresponding steady state is further integrated over one time period. At the instant of impact, the saltation matrix $\mathbf{S}_2$ (Eq. \eqref{eq 30}) is numerically calculated from the $\mathcal{O}(2)$ TDM defined in Eq. \eqref{eq 25} for the impact oscillator. The monodromy matrix is evaluated using Eq. \eqref{eq 32} followed by an eigenvalue analysis to determine the stability of the respective limit cycle for the chosen parameter $\omega$.
\begin{figure}[!]
    \centering
    \includegraphics[scale = 0.8]{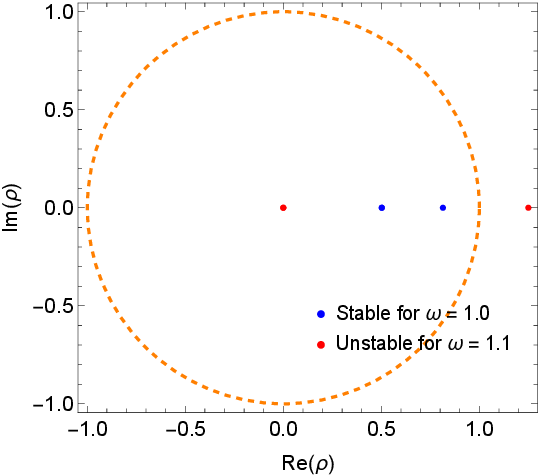}
    \caption{Floquet multipliers for the impact oscillator showing eigenvalues of a stable and an unstable periodic solution corresponding to $\omega = 1.0$ and $\omega = 1.1$ in red and blue respectively. The dashed orange circle of unit radius encloses eigenvalues corresponding to stable periodic solutions.}
    \label{fig 15}
\end{figure}

In Fig. \ref{fig 15}, the eigenvalues of $\mathbf{\Phi}(T)$ are shown in the complex plane for external frequency $\omega$ of $1.0$ and $1.1$, respectively. It is observed that the eigenvalues corresponding to $\omega = 1.0$ are within the unit circle $z = e^{i \theta}$, hence the limit cycle is stable. This inference is also supported by the phase portrait and the Lyapunov spectrum presented in Fig. \ref{fig 10}(a) and (b). However, one of the eigenvalues for $\omega = 1.1$ is outside the unit circle, and thus, the perturbations grow along an eigendirection, leading to divergence in trajectories. The underlying chaotic behaviour can also be observed in Fig. \ref{fig 10}(c) and (d). 

\begin{figure}[!]
    \centering
    \includegraphics[scale = 0.81]{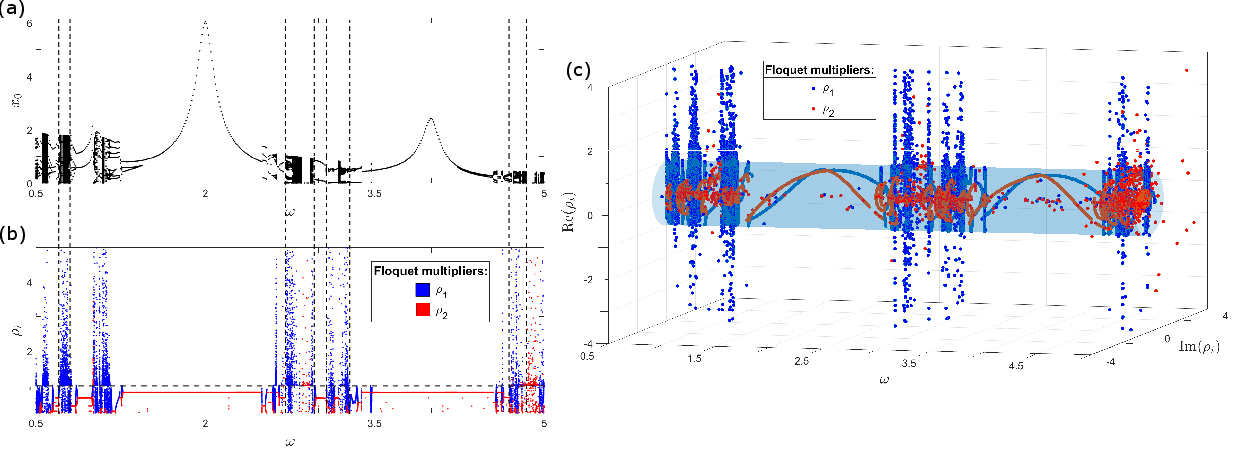}
    \caption{(a) Bifurcation diagram showing amplitude $x_0$ of the impact oscillator vs. external frequency $\omega$. (b) The magnitude of Floquet multipliers $\rho_i$s vs $\omega$. (c) Real and imaginary parts of $\rho_i$s vs $\omega$. Here, $\xi = 0$, $r = 0.8$ and $\omega$ ranges between $0.5 \leq \omega \leq 5.0$..}
    \label{fig 16}
\end{figure}

Figs. \ref{fig 16}(a) and (b) are the bifurcation diagram and the corresponding magnitudes of Floquet multipliers against the driving frequency ranging between $0.5 \leq \omega \leq 5.0$. Dashed vertical lines are shown for comparison between the bifurcation diagram and the Floquet multipliers, highlighting some of the aperiodic chaotic solutions. The dashed horizontal line in Fig. \ref{fig 16}(b) marks $\|\rho\| = 1$ and separates the stable $\omega$ values from the unstable ones. Fig. \ref{fig 16}(c) is a scatter plot of the real and imaginary part of the Floquet multipliers against $\omega$. The blue cylinder with a unit radius is the stable region. All points within this cylinder represent $\omega$ that result in a stable limit cycle, while those lying outside with $\|\rho\| \geq 1$ correspond to unstable parameter values. Note that eigenvalues corresponding to an unstable $\omega$ are scaled down to populate the plot within the defined range for easier visual representation.
%\begin{figure}[!]
%    \centering
%    \includegraphics[scale = 0.25]{Figures/FloquetSpectrum3DImpact.eps}
%    \caption{Floquet multipliers of the impact oscillator in the complex plane against the frequency of the harmonic excitation $0.5 \leq \omega \leq 5.0$ for $r = 0.8$. The blue cylinder of unit radius is the region enclosing Floquet multipliers corresponding to stable periodic solutions.}
%    \label{fig 19}
%\end{figure}

%\subsection{Pair impact oscillator}
Next is the stability analysis of the pair impact oscillator, described by Eq. \eqref{eq 24}. Transients are discarded after $4000$ impacts since there are two impacting surfaces $\Sigma_{2}$ at $\pm \nu/2$. The bifurcation parameter is the forcing amplitude $\alpha$ that drives the cart with frequency $\omega = 1.0$. The time period for a chosen $\alpha$ is evaluated numerically from the recurring time when $\mathbf{y}$ intersects the Poincar\'{e} section $y = 0$. The $\mathcal{O}(2)$ saltation matrix is calculated using the higher-order TDM defined in Eq. \eqref{eq 26} during an impact with either impacting surface. The monodromy matrix $\mathbf{\Phi}$ is evaluated numerically from Eq. \eqref{eq 32} for a chosen $\alpha$, and its eigenvalues determine the stability of the pair-impact oscillator.

\begin{figure}[!]
    \centering
    \includegraphics[scale = 0.8]{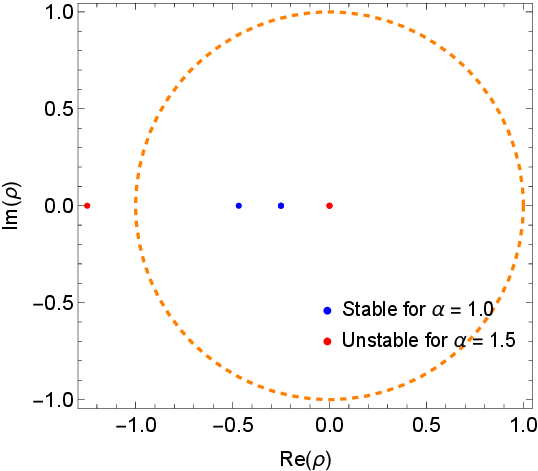}
    \caption{Floquet multipliers for the pair impact oscillator showing eigenvalues of a stable and an unstable periodic solution corresponding to $\alpha = 1.0$ and $\alpha = 1.5$, in blue and red respectively. The dashed orange circle is of unit radius.}
    \label{fig 17}
\end{figure}

In Fig. \ref{fig 17}, the Floquet multipliers are shown for $\alpha = 1.0$ and $1.5$, respectively. The eigenvalues for $\alpha = 1.0$ are within the unit circle, implying the oscillator is stable. This is also confirmed by the observations from the phase portrait and the respective Lyapunov spectrum (Figs. \ref{fig 12}(a) and (b)). However, one of the eigenvalues for $\alpha = 1.5$ is outside the unit circle, and the corresponding limit cycle shows a diverging trajectory. This is also seen in Figs. \ref{fig 12}(c) and (d) where the state space is chaotic and the LLE is positive.

\begin{figure}[!]
    \centering
    \includegraphics[scale = 0.81]{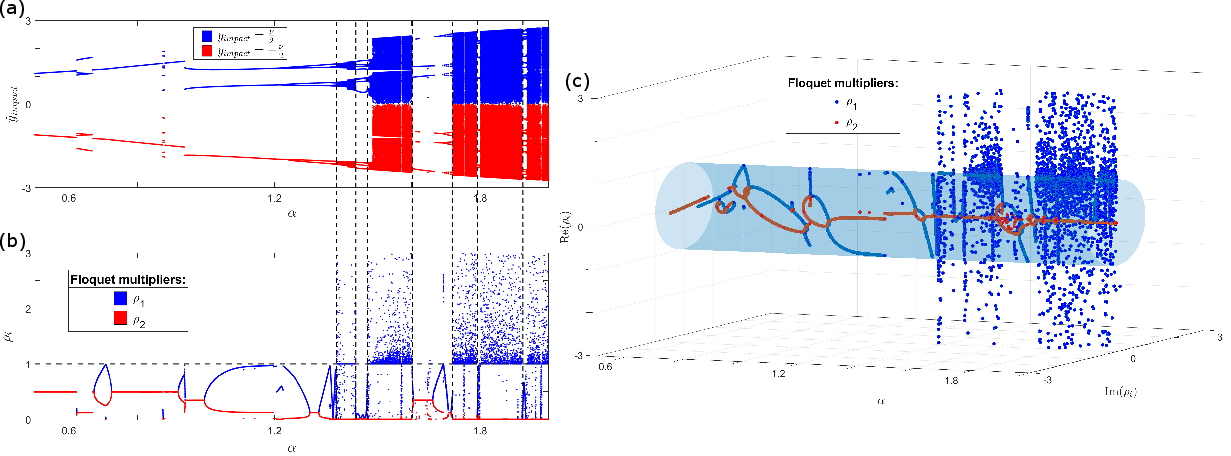}
    \caption{(a) Bifurcation diagram showing velocity $\dot{y}$ of the steady state pair impact oscillator at the instant of impact, \textit{i.e.,} $y = \pm \nu/2$. (b) The corresponding Floquet multipliers $\rho_i$s vs $\alpha$. (c) Real and imaginary parts of $\rho_i$s vs. $\alpha$. Here, $\alpha$ ranges between $0.5 \leq \alpha \leq 2.0$, $\omega = 1.0$ and $r = 0.7$.}
    \label{fig 18}
\end{figure}

In Fig. \ref{fig 18}(b), the magnitude of Floquet multipliers for the pair impact oscillator against the bifurcation parameter $\alpha$ is shown. A bifurcation diagram of the velocity $\dot{y}$ at the instant of impact is shown against $\alpha$ for reference in Fig. \ref{fig 18}(a). The dashed line where $\|\rho\| = 1$ separates the stable periodic limit cycles from the unstable diverging trajectories. $\|\rho\| \geq 1$ corresponds to values of $\alpha$ for which trajectories near the periodic limit cycle diverge according to Floquet theory. Fig. \ref{fig 18}(c) shows the real and imaginary parts of the Floquet multiplier $\rho_i$ for different values of $\alpha$. Eigenvalues that lie outside the cylinder with unit radius correspond to diverging trajectories for the respective $\alpha$. Such eigenvalues with $\|\rho\| >> 1$ are scaled down for better visual representation. 
%\begin{figure}[!]
%    \centering
%    \includegraphics[scale = 0.25]{Figures/FloquetSpectrum3DPair.eps}
%    \caption{Floquet multipliers of the pair impact oscillator in the complex plane for $0.5 \leq \alpha \leq 2.0$, $\omega = 1.0$ and $r = 0.7$. The blue cylinder of unit radius is the region enclosing Floquet multipliers corresponding to stable periodic solutions for the respective $\alpha$.}
%    \label{fig 22}
%\end{figure}

\section{Stability analysis during period-adding cascades}\label{sec 7}

This section implements the higher-order saltation matrix and the $\mathcal{O}(2)$ TDM to predict the bifurcation behaviour in impact oscillators where period-adding cascades are observed. The regime of period-adding cascades is chosen to test whether the characteristic exponents obtained from the higher-order theory can show how orbits approach a critical parameter value, after which a DIB is observed during direct numerical simulations. Hence, an impact oscillator is considered for this while varying the barrier distance $\sigma$ corresponding to $\xi = 2.0$, $\omega = 2.0$ and $r = 0.8$. During an impact at $t_i$, the higher-order TDM is given by Eq. \eqref{eq 34}. Note that the TDM defined in Eq. \eqref{eq 34} depends on $\xi$, $\omega$ and $t_i$ that is not observed in the linearized saltation matrix.
\begin{align} \label{eq 34}
    \mathbf{y}_+ =  \begin{bmatrix}
                    \bigg( y_1 + \delta^{\xi}_+ v(1+r) + \delta^{\xi}_+ y_2(1-r^2) + \dfrac{(\delta^{\xi}_+)^2}{2}(-\sigma + \cos{(\omega t_i)})(1 - r - 2r^2) - v \xi (\delta^{\xi}_+)^2 (1 - r^2) \bigg) \\
                   \bigg( -r y_2 - r \delta^{\xi}_+(-\sigma - 2\xi v + \cos{(\omega t_i)}) - \delta^{\xi}_+(-\sigma + 2 r \xi v + \cos{(\omega t_i)})+  \\
                    r \delta^{\xi}_+(y_1 + 2y_2 \xi) + \dfrac{r(\delta^{\xi}_+)^2}{2}(v + 2\xi(-\sigma - 2\xi v + \cos{(\omega t_i)}) + \omega \sin{(\omega t_i)})+  \\
                    \delta^{\xi}_+(y_1 + 2r^2 y_2 \xi) + (\delta^{\xi}_+)^{2}(v + 2r^2 \xi(-\sigma - 2\xi v + \cos{(\omega t_i)}) + \omega \sin{(\omega t_i)})+  \\ 
                     \dfrac{(\delta^{\xi}_+)^2}{2}(r v + 2r\xi(-\sigma + 2 r \xi v + \cos{(\omega t_i)}) - \omega\sin{(\omega t_i)}) \bigg)
                  \end{bmatrix}_{2\times 1}
\end{align}

\begin{figure}[!]
    \centering
    \includegraphics[scale = 0.9]{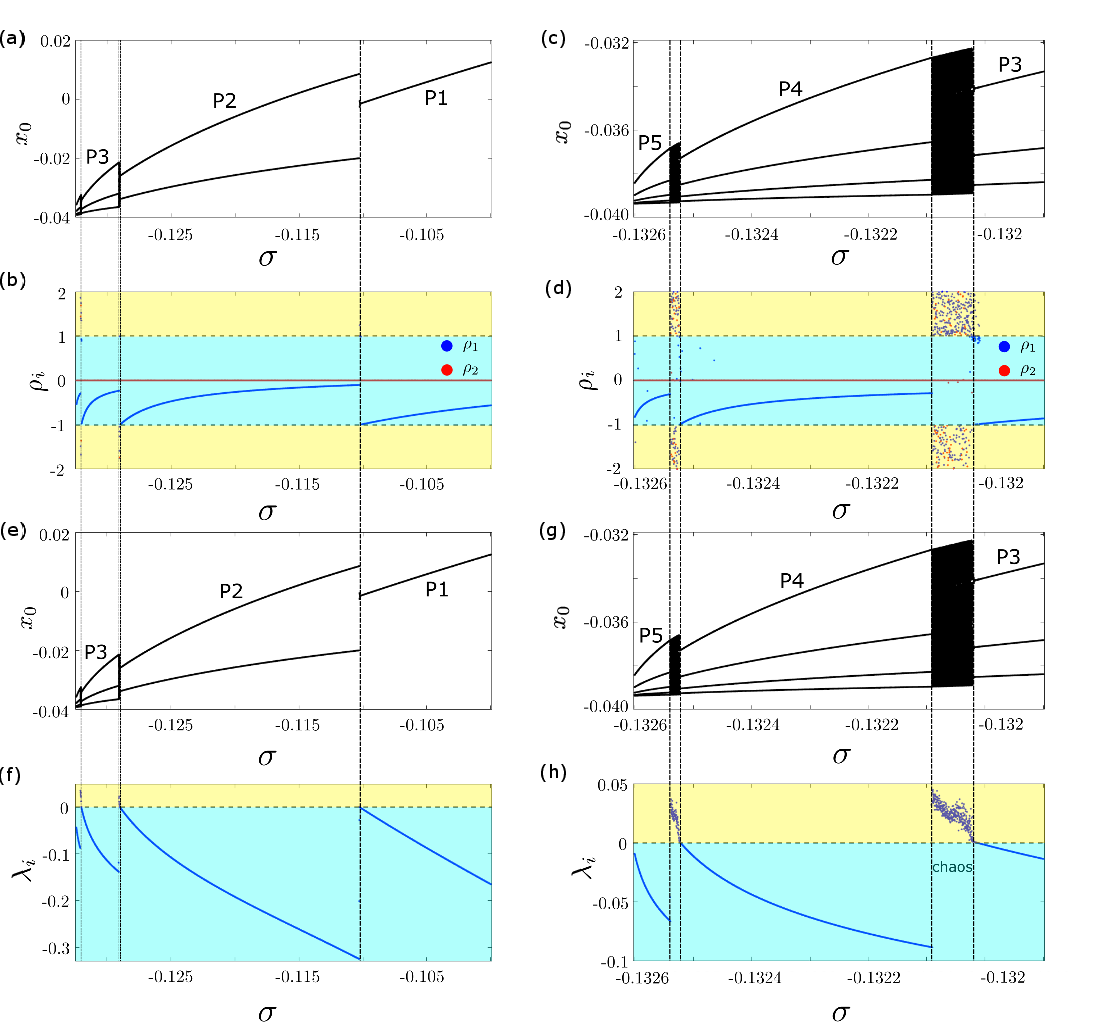}
    \caption{Bifurcation diagram showing $x_0$ versus barrier distance $\sigma$ ranging between $-0.1324 \leq \sigma \leq -0.10$ in (a) and (e) and $-0.1326 \leq \sigma \leq -0.1319$ in (c) and (g). The corresponding magnitude of Floquet multipliers and LEs vs $\sigma$ ranging between $-0.1324 \leq \sigma \leq -0.10$ in (b) and (f) and $-0.1326 \leq \sigma \leq -0.1319$ in (d) and (h). The bounded region $-1 \leq \rho_i \leq 1$ and $\lambda_i \leq 0$ is the stable region shown in cyan, while the unstable region is the yellow shaded area.}
    \label{fig 19}
\end{figure}

Figs. \ref{fig 19}(b) and (d) shows the magnitude of Floquet multipliers $\rho_1$ and $\rho_2$ as the barrier distance is varied between $-0.1324 \leq \sigma \leq -0.10$ and $-0.1326 \leq \sigma \leq -0.1319$. A bifurcation diagram showing amplitude response $x_0$ versus $\sigma$ is given for reference in Figs. \ref{fig 19}(a) and (c). The Floquet multipliers were obtained using the higher-order saltation matrix and the TDM defined in Eqs. \eqref{eq 30} and \eqref{eq 34}. Period-adding cascades of solutions separated by chaotic orbits were observed here with periodicity P$\#$ being defined as the number of intersections with a Poincar\'e section $\Sigma_1: x = 0$ with $\ddot{x} < 0$. Results show that Floquet multipliers have a norm less than unity for the chosen $\sigma$ corresponding to a stable orbit (cyan region). As $\sigma$ is varied, the largest Floquet multiplier approaches $-1$ and becomes unstable (yellow region). This indicates that the corresponding limit cycle becomes unstable via a period-doubling bifurcation, and immediately, a new periodic orbit is observed. This is verified in the bifurcation diagram, where a DIB occurs in the same value of $\sigma$; see Figs. \ref{fig 19}(a) and (c). However, the occurrence of chaotic solutions cannot be determined using Floquet multipliers, as they are applicable to dynamical systems where the Jacobian matrix is periodic. To validate the chaotic orbits observed in the amplitude response of Fig. \ref{fig 19}(c), the largest LE (LLE) is estimated in Figs. \ref{fig 19}(f) and (h). Results show that the LLE is always negative for $\sigma$ values with a stable periodic orbit. As $\sigma$ is varied, the LLE approaches zero, and a DIB is observed in the amplitude response. Additionally, the LLE in Fig. \ref{fig 19}(h) is positive for a range of $\sigma$s, implying the occurrences of chaotic orbits between period three and period four solutions; see Fig. \ref{fig 19}(g). A comparison with the bifurcation diagram verifies the result. Therefore, the algorithms of \ref{algo 1} and \ref{algo 2}, which implement a higher-order TDM and saltation matrix, can correctly predict bifurcation behaviour for hybrid dynamical systems. Note that there is minimal change in the computational cost when replacing the first-order with the higher-order TDM. However, the computational cost is drastically reduced when the fixed points are obtained analytically, followed by the implementation of the higher-order TDM for stability analysis \cite{chawla2025higher}. Further, the derivation of the proposed higher-order TDM is generalised for any $\mathbf{x} \in \mathbb{R}^n$ where $n \in \mathbb{Z}_+$. The methodology provides a framework for identifying the governing factors and their higher-order correction terms, which significantly enhances the accuracy of post-impact state predictions in hard-impact oscillators with $n > 2$ \cite{chawla2024wake, chawla2023higher, chawla2024discontinuity}. In particular, incorporating the higher-order flight time yields a substantial refinement over first-order approaches, such as those relying solely on saltation matrices. This methodology can be naturally extended to piecewise-smooth dynamical systems with different degrees of smoothness (DOS), such as Fillipov systems with DOS 1 or piecewise-continuous systems with DOS 2. We have developed a corresponding higher-order TDM for Filippov systems based on this same principle; the details are available in \cite{chawla2025higherFilippov}.       

\section{Conclusions} \label{sec 8}
\noindent For a hybrid dynamical system of order $\mathbb{R}^{n}$, a closed-form expression of the transverse discontinuity mapping with higher-order corrections has been derived. In addition to discrepancies near the degenerate grazing conditions, the work highlights how, even under non-degenerate conditions, the discontinuity mapping can inaccurately deduce the state of stability of the system. The proposed higher-order corrections to the time of flight presented in this paper address this issue by avoiding incorrect identification of an impact. This is deduced from the derived quadratic equation that analytically proves that not all orbits initiated in the local neighbourhood of an impacting state reach the barrier. When perturbations become too large, the discriminant of the higher-order flight time becomes negative. This indicates that perturbed orbits continue evolving in the phase-space away from the barrier. First-order transverse discontinuity mapping and the saltation matrix cannot capture this since they predict that all orbits reach the barrier. Additionally, the first-order flight time inherently possesses singularity during low-velocity impacts or grazing incidence. The higher-order correction terms resolve this limitation by ensuring that the discontinuity mappings do not diverge for low-velocity impacts. 

This retention of higher order terms also improves the estimates of the mapped states of the system, which has implications for the stability of the orbit under consideration. The problem of overestimation of mapped states on the discontinuity boundary while using the first-order saltation matrix is addressed. The proposed correction terms incorporate system parameters and the functional form of the driving force of the hybrid system under consideration, leading to much closer estimates to numerical simulations. This explicit dependence on crucial parameters like the driving force is absent in a linearized approximation. The proposed approach is demonstrated using two examples of hybrid impacting systems, with single and multiple impacting barriers, respectively. It is shown that the semi-analytically obtained higher-order saltation matrix, derived with the variational approach, is able to better predict the states of the orbits post saltations, as well as provide more accurate stability estimates. %Thus, it is quite clear that the linearized approximation of TDM incorrectly predict an impact, and is not enough to reveal the dynamical behaviour of the original system as demonstrated by numerical simulations. 
%Moreover, results implementing the higher-order saltation terms are applicable when the initial separation between trajectories is such that it leads to a large difference in impact times.
%orders of magnitude larger than the ones for which the linearized formalism is effective. 
%A higher periodic or aperiodic trajectory would be effectively analytically mapped with the help of the proposed formalism.
% The higher-order formulation proposed in this paper can be used to obtain insights to the stability estimates of hybrid systems. This is demonstrated with two hybrid systems possessing single and multiple undeformable barriers as constraints to motion.
The algorithm used to compute stability estimates using higher-order TDM to evaluate the LE spectra, along with its implementation method for the above examples, is provided. The analysis does not require any prior knowledge of the dynamical states at the instances of impact, unlike the approaches used for transcendental maps. %With driving force frequency as the bifurcation parameter, bifurcation diagrams are obtained to corroborate stability estimates obtained for underlying attractors. The Lyapunov spectrum is observed to be consistent with the results obtained from bifurcation analysis, even for low-velocity impacts near the bifurcation boundaries. %It was observed that the impact oscillators exhibit chaos as indicated by positive largest Lyapunov exponents.
% A numerical approach to obtain the saltation matrix at the discontinuity boundary with higher-order corrections was proposed where a numerical matrix inversion accommodated second-order terms within the matrix form. This method can be extended to accommodate even higher-order terms if necessary. Monodromy matrices for the hybrid systems under consideration are computed, and an eigenvalue analysis provides the respective Floquet multipliers. The dynamical behaviour observed via examination of the Floquet multipliers and the bifurcation diagram are in agreement with each other. 
 
The results from this work are directly applicable to the study of complex higher periodic or aperiodic orbits in hybrid systems of order $\mathbb{R}^n$, where analytically accurate estimates for trajectories interacting with the discontinuity boundary and quantification of stability are obtained, especially near bifurcation regimes when one of the attractors loses stability and/or new stable states are born. 
%A further extension of this work to Filippov systems is being taken up in a separate study.

\section*{Declarations}
The authors acknowledge the funding of Research Ireland 22FFP-P11457 HarMonI, along with NexSys 21/SPP/3756 and  RC2302-2 MaREI, and Sustainable Energy Authority of Ireland funded RDD/604 TwinFarm, RDD/966 FlowDyn.  

\section*{Conflict of interest}
The authors declare that they have no conflict of interest.

\section*{Availability of data}
Not applicable.

\section*{Availability of code}
All implemented codes are available upon request to the corresponding author.

\appendix

\section{Lyapunov exponents using higher-order transverse discontinuity mapping}\label{app a}

\renewcommand{\thealgorithm}{A.\arabic{algorithm}}
\setcounter{algorithm}{0}

For a dynamical system of order $\mathbb{R}^n$, the exponential divergence along the $i^{th}$ orthogonal eigenvector can be found using,
\begin{equation} \label{eq a1}
    \lambda_i = \dfrac{1}{\tau N} \Sigma_{n = 1}^N \log{\Big(\dfrac{r_n}{r_0}\Big)}.
\end{equation}

The $i^{th}$ LE in Eq. \eqref{eq a1} is evaluated by measuring the growth of the variation in the $i^{\mathrm{th}}$ direction after every time period $\tau = 2\pi/\omega$ {\it i.e.,} in a stroboscopic fashion. Since the state is of the order $\mathbb{R}^n$, there are $n$ independent solutions of the corresponding variational form. Therefore, any arbitrary solution of the perturbed orbit can be decomposed along these $n$ eigenvectors. To measure the $i^{th}$ LE along any of these eigenvectors, a QR decomposition (QRD) is carried out using the Gram-Schmidt process. The process yields $n$ orthogonal perturbed vectors, which can be encapsulated in a hypersphere of dimension $n$. The growth or decay of this hypersphere along the trajectory over time is an indicator of the stability of the dynamical system. The LE along an eigendirection is calculated by numerically integrating each of these orthogonal vectors and measuring the change in magnitude of these vectors after a time interval of $\tau$ where $r_0$ and $r_n$ in Eq. \eqref{eq a1} are the initial and final magnitudes of the perturbed vector after elapsed time $\tau$. However, for chaotic systems, the perturbed vectors might quickly diverge from the actual trajectory, and the linearized variational form might not be able to capture the actual dynamics of the perturbed trajectory. Therefore, the initial perturbed hypersphere is kept small by using a scaling factor $r_0$ in Eq. \eqref{eq a1}. Furthermore, to obtain accurate values of LE, $\lambda_i$ in Eq. \eqref{eq a1} has been averaged out over several computations. The algorithm for estimating LEs by incorporating a higher-order TDM is presented in Algorithm \ref{algo 1}.
\begin{algorithm}[!]
    \caption{Lyapunov exponent for hybrid systems using TDM} \label{algo 1}
    \begin{algorithmic}
    \State 1. Initialize: $\mathbf{x}(0)$ ensuring $H(\mathbf{x} \geq 0)$ and $\omega$ or $\alpha$ \Comment{Bifurcation parameter} 
    \State 2. Initialize: $\mathbf{y}_i(0)$ for $i \leq n$ using QRD 
    \State 3. Rescale: $\mathbf{y}_i(0) \gets r_0\times \mathbf{y}_i(0)$ and set $n_{max}$ \Comment{Maximum allowable impacts}
    \While{count $\leq n_{max}$}
    \State Integrate: $\dot{\mathbf{x}} = \mathbf{F}(\mathbf{x})$
    \If{$H(\mathbf{x}) = 0$} \Comment{Occurrence of impact}
        \State Evaluate: $\delta^{\xi}_+$, $\mathbf{R}(\mathbf{x})$ and $\mathbf{y}_+$
        \State Apply reset maps: $\mathbf{x} \gets \mathbf{R}(\mathbf{x})$ and $\mathbf{y}_i \gets \mathbf{y}_{i,+}$ \Comment{Implement $\mathcal{O}(2)$ TDM} 
    \EndIf
    \If{$\big( t \% \dfrac{2\pi}{\omega} \big) = 0$}
        \If{count $\geq n_{max}/2$}
            \State Store: $r_i \gets \dfrac{1}{r_0} \| \mathbf{y}_i \|$ \Comment{store the growth rate}
        \EndIf
    \State Reinitialize: $\mathbf{y}_i \gets r_0 \times \text{QRD of } \mathbf{y}_i$ 
    \EndIf
    \EndWhile
    \State 4: Evaluate: $\log_e r_i$ \Comment{Store all $log_e r_i$}
    \State 5: LE$_i \gets \dfrac{\omega}{2\pi} \times \text{Partial sum of } \log_e r_i$
    \State 6: $\lambda_i \gets \langle \text{LE}_i \rangle$ \Comment{Mean of all LEs}
    \end{algorithmic}
\end{algorithm}

\section{Floquet multipliers using numerically obtained higher-order saltation matrix}\label{app b}

\renewcommand{\thealgorithm}{B.\arabic{algorithm}}
\setcounter{algorithm}{0}

Floquet theory \cite{floquet1883linear} dictates that eigenvalues of the monodromy matrix determine the stability of a limit cycle. The eigenvalues or characteristic multipliers \cite{nayfeh2008applied} determine the behaviour of orbits in the local linear neighbourhood of steady states. However, for hybrid dynamical systems, the monodromy matrix cannot be directly evaluated by integrating the variational equation with initial conditions corresponding to an orthogonal basis. This is due to discontinuities occurring in the state space near the discontinuity boundary. The higher-order saltation matrix is rectified by defining a state transition between the discrete mappings. Algorithm \ref{algo 2} outlines the method to evaluate the Floquet multipliers using numerically obtained saltation matrices comprising higher-order correction terms.

\begin{algorithm}[!]
    \caption{Floquet multipliers from monodromy matrix $\mathbf{\Phi}$} \label{algo 2}
    \begin{algorithmic}
    \State 1. Initialize: $\mathbf{x}(0)$ ensuring $H(\mathbf{x}) \geq 0$ and $\omega$ or $\alpha$ \Comment{Bifurcation parameter}
    \State 2. Initialize: $\mathbf{\Phi} = \textit{I}_{n \times n}$, T$ = 100$, count $= 0$, $n_{max}$ \Comment{Maximum allowable impacts}
    \While{count $\leq$ $n_{max}$}
    \State Integrate: $\dot{\mathbf{x}} = \mathbf{F}(\mathbf{x})$
    \If{$H(\mathbf{x}) = 0$} \Comment{Occurrence of impact}
        \State Reset Map: $\mathbf{x} \gets \mathbf{R}(\mathbf{x})$
    \EndIf
    \If{$\dot{x} = 0$ $\&\&$ count $\geq n_{max}/2$}
        \State $T = \{t \in \mathbb{R}^1:x(t + \tilde{n}T) = x(t), \tilde{n}\in \mathbb{I} \}$ \Comment{Store $x$, $t$ for evaluation of time period $T$}
    \EndIf
    \If{count $=$ $n_{max}$ - 100} \Comment{Remove transients}
        \State Store: $\mathbf{x}_{init}$ $\gets \mathbf{x}(t)$ and $t_{init} \gets t$ 
    \EndIf
    \EndWhile
    \State Initialize: $\mathbf{x} \gets \mathbf{x}_{init}$ at $t_{init}$ \Comment{Begin at steady state} 
    \State Initialize: $r_0 \ll 1$, $\mathbf{Y}_1(0) = r_0 \textit{I}_{n \times n}$, $\mathbf{Y}_2(0) = \textit{I}_{n \times n}$ \Comment{Evaluation of $\mathbf{S}_2$ and $\mathbf{\Phi}_i$}
    \While{$t_{init} \leq t \leq (t_{init} + T)$} \Comment{Integrate over period $T$}
        \State Integrate: $\dot{\mathbf{x}} = \mathbf{F}(\mathbf{x})$, $\dot{\mathbf{Y}_1} = (\nabla \mathbf{F})^T\cdot\mathbf{Y}_1$ and $\dot{\mathbf{Y}_2} = (\nabla \mathbf{F})^T\cdot\mathbf{Y}_2$ 
        \If{$H(\mathbf{x}) = 0$} \Comment{Occurrence of impact}
        \State $\mathbf{\Phi}_i \gets \mathbf{Y}_2$ and $\mathbf{\Phi} \gets \mathbf{\Phi}_i\cdot\mathbf{\Phi}$
        \State Evaluate $\delta$ and $\mathbf{y}_+$ from $\mathbf{x}_4$ for all $\mathbf{y}_i$ \Comment{TDM}
        \State Evaluate $\mathbf{Y}_{+, impact}$ from $\mathbf{y}_+$ and $\mathbf{Y}_{-\text{, impact}} \gets \mathbf{Y}_1$
        \State Evaluate $\mathbf{S}_2 \gets \mathbf{Y}_{+, impact}\cdot\mathbf{Y}_{-, impact}^{-1}$ and set $\mathbf{\Phi} \gets \mathbf{S}_2 \cdot \mathbf{\Phi}$ \Comment{saltation matrix}
        \State Reset Map: $\mathbf{x} \gets \mathbf{R}(\mathbf{x})$
        \State Reinitialize: $\mathbf{Y}_1 \gets r_0 \textit{I}_{n \times n}$ and $\mathbf{Y}_2 \gets \textit{I}_{n \times n}$
        \EndIf
    \EndWhile
    \State 3. $\mathbf{\Phi} \gets \mathbf{Y}_2\cdot\mathbf{\Phi}$ \Comment{Evaluate monodromy matrix}
    \State 4. Evaluate eigenvalues of $\mathbf{\Phi}$ to get Floquet multipliers
    \end{algorithmic}
\end{algorithm}

\bibliographystyle{elsarticle-num-names}
\bibliography{References}

\end{document}